\documentclass[fleqn,usenatbib]{mnras}

\usepackage{newtxtext,newtxmath}
 
\usepackage[T1]{fontenc}

\DeclareRobustCommand{\VAN}[3]{#2}
\let\VANthebibliography\thebibliography
\def\thebibliography{\DeclareRobustCommand{\VAN}[3]{##3}\VANthebibliography}

\usepackage{graphicx}	
\usepackage{amsmath}	
\usepackage[version=4]{mhchem}
\usepackage{epstopdf}

\begingroup
\catcode`\_=\active
\gdef_#1{\ensuremath{\sb{\mathrm{#1}}}}
\endgroup
\mathcode`\_=\string"8000
\catcode`\_=12

\newcommand{\cm}[1]{cm$^{#1}$}                                  
\newcommand{\htwo}{H$_2$}                            
\newcommand{\hcc}{~H~\cm{-3}}                        
\newcommand{\msun}{~\mbox{M}$_{\odot}$}                      
\newcommand{\pc}{~\mbox{pc}}                         
\newcommand{\pch}{~\mbox{pc/h}}                      

\newcommand{\jj}{\,J$_{21}$}                         
\newcommand{\jlw}{\,J$_{\mbox{\scriptsize{LW}},21}$}    
\newcommand{\jxz}{\,J$_{\mbox{\scriptsize{X0}},21}$}   

\newcommand{\nh}{n_{\mbox{\scriptsize{H}}}}        

\newcommand{\ramses}{\texttt{RAMSES}}                

\def\hide#1{}

\title[First stars in an X-ray background]{Population~III Star Formation in an X-ray background: II.\\ Protostellar Discs, Multiplicity and Mass Function of the Stars}

\author[J. Park, M. Ricotti and K. Sugimura]{
Jongwon Park,$^{1}$
Massimo Ricotti,$^{1}$
and Kazuyuki Sugimura$^{1,2}$
\\
$^{1}$Department of Astronomy, University of Maryland, College Park, MD 20742, USA\\
$^{2}$Astronomical Institute, Graduate School of Science, Tohoku University, Aoba, Sendai 980-8578, Japan\\
}

\date{Accepted XXX. Received YYY; in original form ZZZ}

\pubyear{2021}

\begin{document}
\label{firstpage}
\pagerange{\pageref{firstpage}--\pageref{lastpage}}
\maketitle

\begin{abstract}
Disc fragmentation plays an important role in determining the number of primordial stars (Pop~III stars), their masses, and hence the initial mass function. In this second paper of a series, we explore the effect of uniform FUV \htwo-photodissociating and X-ray radiation backgrounds on the formation of Pop~III stars using a grid of high-resolution zoom-in simulations. We find that, in an X-ray background, protostellar discs have lower surface density and higher Toomre $Q$ parameter, so they are more stable. For this reason, X-ray irradiated discs undergo fewer fragmentations and typically produce either binary systems or low-multiplicity systems. In contrast, the cases with weak or no X-ray irradiation produce systems with a typical multiplicity of $6\pm 3$. In addition, the most massive protostar in each system is smaller by roughly a factor of two when the disc is irradiated by X-rays, due to lower accretion rate. With these two effects combined, the initial mass function of fragments becomes more top-heavy in a strong X-ray background and is well described by a power-law with slope $1.53$ and high-mass cutoff of $61$\msun. Without X-rays, we find a slope $0.49$ and cutoff mass of $229$\msun. Finally, protostars migrate outward after their formation due to the accretion of high-angular momentum gas from outside and the migration is more frequent and significant in absence of X-ray irradiation.
\end{abstract}





\section{Introduction}

In the last 20 years theoretical progress has been made to better understand the formation of the first zero-metallicity stars (Pop~III) that formed in the universe  \citep{omukai1998,bromm2001,abel2002,yoshida2008,turk2009,clark2011,hosokawa2011,sugimura2020}. However, predictions on the number and masses of the Pop~III stars are still very uncertain. Similarly uncertain are the predictions on the number of pair-instability SNe (PISNe) and hypernovae from Pop~III stars and the number of IMBHs they produce. In the era of gravitational wave astronomy \citep{abbott2016}, the census of intermediate mass black holes (IMBHs) detected in binary systems is bound to improve, and next generation optical and IR space telescopes (JWST and Roman space telescope) promise to directly detect the first light from Pop~III star clusters, PISNe and hypernovae they may produce \citep{whalen2014}. It is therefore important to refine our models and predictions for the  the rate of formation of the first stars and their remnants. 

Part of the difficult in making these predictions are global feedback processes, in particular the Lyman-Warner (LW) H$_2$ dissociating background and the X-ray radiation background, regulating their formation in small mass haloes (minihaloes) and their initial mass function \citep[][]{oh2001, venkatesan2001, machacek2003, RicottiO:04, RicottiOG:05, jeon2014, xu2016}. The LW radiation is emitted directly by Pop~III stars and second generation stars, while high-mass X-ray binaries (HMXBs), accreating IMBHs, and supernova/hypernova explosions are sources of an X-ray radiation background associated with Pop~III star formation \citep{xu2014, jeon2014, jeon2015, ricotti2016, xu2016}.
The LW radiation background has always a negative feedback on the formation of Pop~III stars as it dissociates H$_2$, suppressing cooling of pristine gas. The X-ray background can both suppress Pop~III star formation in the smallest minihaloes due to intergalactic medium (IGM) heating (increasing the Jeans mass in the IGM) and promote Pop~III star formation by increasing the electron fraction of gas collapsed into minihaloes and thus promoting H$_2$ formation via the catalyst H$^-$.
The number of Pop~III stars depends on the minimum dark matter halo
mass in which a zero-metallicity star can form as a function of redshift (the smaller the critical mass the more numerous the minihaloes hosting Pop~III stars) and on the multiplicity of stars per minihalo \citep{hirano2014,susa2014,stacy2016}. Since both the critical minihalo mass and the multiplicity of stars depend on the radiation backgrounds produced by Pop~III stars, a feedback loop is at play.
An simple analytic model of this feedback loop was presented in \citet{ricotti2016}, in which the number of Pop~III stars in the early universe and the radiation background they produce was estimated self-consistently. He finds that X-rays emitted during the unavoidable death of Pop~III stars in cosmic supernova or hypernova explosions, can significantly increase the number of Pop~III stars forming in the universe with respect to models without X-rays. However, if the X-ray emissivity per minihalo is too large, for instance because of significant contributions from HMXBs or accreting IMBHs from Pop~III stars or miniquasars from direct collapse BHs, X-rays can have a negative feedback effect by heating excessively the IGM. 

The buildup and cosmological effects of an X-ray background, including its effect on the formation of the first stars, have been studied by many authors \citep{oh2001, venkatesan2001, machacek2003,inayoshi2011,jeon2014,xu2014,xu2016}. However, its impact on the properties of Pop~III stars has receive less attention. The pioneering work of \citet{hummel2015} focused on the initial gas collapse, disc fragmentation, and multiplicity of Pop~III stars in presence of an X-ray background using cosmological simulations. Evolving their simulations for the first 5,000~yrs after the formation of the first sink particle, they concluded that the impact of X-rays on Pop~III star formation is minimal. However, the process of accretion onto protostellar cores takes typically $\sim 10^4 - 10^5$ yrs \citep{mckee2008,hosokawa2011,sugimura2020}, hence these simulations may have been evolved for an insufficient length of time to fully capture the evolution of Pop~III stars and protostellar discs.


In \citet*[][hereafter Paper~I]{ParkRS:21a}, we describe a set of zoom cosmological simulations of the formation of Pop~III stars in three minihaloes with different masses and growth histories, irradiated by a range of intensities of the LW and and X-ray background. The intensity of the background is constant as a function of time in physical units and has a power-law shape with flat slope for the LW radiation and slope $1.5$ for the X-ray radiation. We explore a grid of $7\times7=49$ different combinations of LW and X-ray backgrounds: \jlw$=0,10^{-3}, 0.01,0.1,1,10,100$, \jxz$=0,10^{-6},10^{-5}, 10^{-4}, 10^{-3},0.01,0.1$, where the intensities are in units of $10^{-21}$~erg~s$^{-1}$~\cm{-2}~Hz$^{-1}$~sr$^{-1}$.

In Paper~I, in agreement with \citet{ricotti2016}, we find that the X-ray radiation background generally promotes the initial gas collapse in small mass minihaloes, while the LW background delays it by regulating the amount of \htwo\ formation. However, if the X-ray background is too intense, gas heating suppresses Pop~III star formation in haloes with virial temperature $T_{vir}<T_{gas}$. Below this minihalo mass-dependent threshold, enhancement of the \htwo\ abundance produced by X-ray ionization of the gas, reduces the critical mass above which a minihalo can host a Pop~III star to $\sim 10^5$\msun.  The positive feedback effect of X-rays is most important when it offsets the negative feedback of an intense \htwo-dissociating LW radiation background. In this case the critical mass can be reduced by a factor of ten. Hence, X-ray radiation can increase the number of minihaloes forming Pop~III stars in the early Universe by about the same factor.
\begin{figure}
    \centering
	\includegraphics[width=0.48\textwidth]{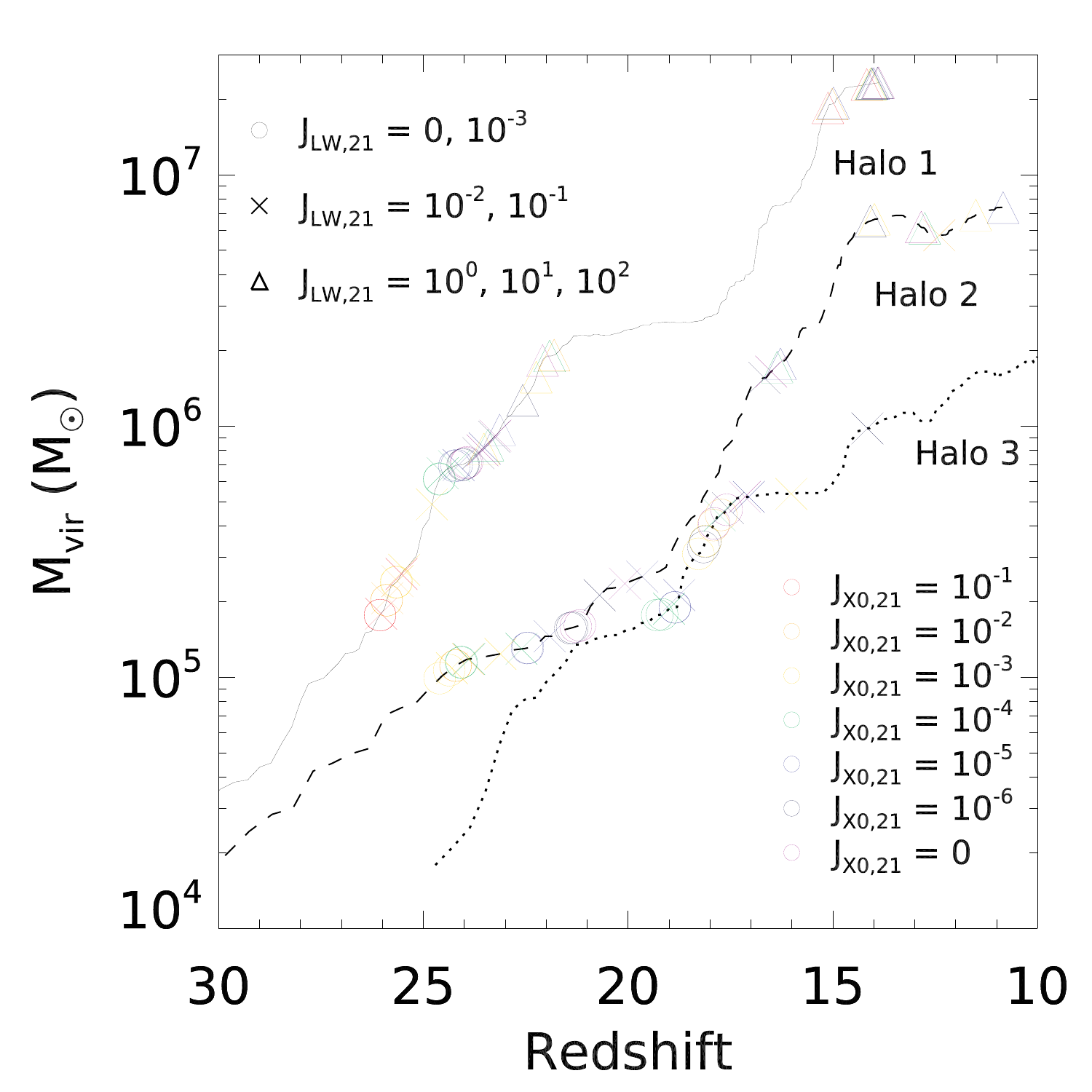}
    \caption{Halo mass as a function of redshift. The virial masses of Halo~1, Halo~2 and Halo~3 are shown with different lines. The positions of symbols refer to the redshift of the formation of Pop~III stars and the masses of their host minihalos. Different symbols and colors indicate the intensity of LW and X-ray backgrounds as indicated by the legend.}
    \label{fig:halo_growth}
\end{figure}
We also found that total mass in Pop~III stars in each minihalo is lower in a sufficiently intense X-ray background. X-ray irradiation produces a net cooling effect on the collapsing protostellar core by increasing the H$_2$ fraction. Efficient gas cooling reduces the gas sound speed and consequently the accretion rate on collapsing protostellar core, and its final mass.

In this work, the second in a series, we investigate how the X-ray and LW radiation backgrounds affect the protostellar disc properties, binarity/multiplicity of Pop~III stars, including their separation and dynamics, and the initial mass function of Pop~III stars.
The paper is organized as follows. In Section~\ref{sec:sim} we briefly summarize the simulations. We discuss the properties of circumstellar discs in Section~\ref{sec:frag} and the multiplicity of Pop~III stars in Section~\ref{sec:mult}. In Section \ref{sec:disc} we provide a summary and discussion.

\section{Simulations and Methods}
\label{sec:sim}
The simulations and the included physical processes are described in detail in Paper~I. Here we summarize the main aspects of the simulations for the sake of completeness. We use the adaptive mesh refinement (AMR) cosmological code \ramses\ \citep{teyssier2002}, with radiation transfer \citep{rosdahl2013}. We perform zoom-in cosmological simulations on three haloes: the mass growth history of these three haloes is shown in Figure~\ref{fig:halo_growth} along with symbols showing the redshift of formation of the Pop~III stars under different intensities of an externally imposed X-ray background.

Inside each zoom-in region, the mass of DM particles is $\sim 800$ \msun, and the cell refinement criteria are: i) Lagrangian (cells must contain less than 8 DM particles) and, ii) Jeans refinement criteria (Jeans length is resolved with at least $N_{J}$ cells). For the latter condition, for cell sizes smaller than $\sim 1 \pch$ comoving, we adopt $N_{J}=16$ in order to prevent any possible artificial fragmentation and better resolve possible turbulent motions. If the size of a cell is greater than $\sim 30 \pch$ (comoving), we adopt $N_{J}=4$ to save computational time. Any cells between these two scales are refined with $N_{J}=8$. The size of a smallest cell is $0.00375 \pch$ (comoving). At $z=20$, this corresponds to a physical size of $2.63 \times 10^{-4} \pc$ (or $54$ au). The corresponding AMR levels are shown in Table~\ref{tab:sim}. The initial conditions of the DM-only and zoom-in simulations are generated with MUSIC \citep{hahn2011}. The assumed cosmological parameters are $h=0.674, \Omega_{m}=0.315, \Omega_{\Lambda}=0.685, \Omega_{b}=0.0493, \sigma_8=0.811$ and $n_s=0.965$ \citep{planck2018}.
\begin{table}
	\centering
	\caption{Summary of the simulations.}
	\label{tab:sim}
	\begin{tabular}{ | r | c | c | c | c | }
		\hline
        & $M_{\mbox{\scriptsize{Vir}}}$ (z=15.7) & $M_{\mbox{\scriptsize{DM}}}$ (zoom-in) & Box size & $l_{\mbox{\scriptsize{max}}}$ \\
		\hline
        Halo~1 & $7.9\times10^6$ \msun & 800 \msun & 1 Mpc/h$^3$ & 28 \\
		\hline
        Halo~2 & $4.4\times10^6$ \msun & 800 \msun & 2 Mpc/h$^3$ & 29 \\
		\hline
        Halo~3 & $7.0\times10^5$ \msun & 800 \msun & 1 Mpc/h$^3$ & 28 \\
        \hline
	\end{tabular}
\end{table}

We consider all important chemical reactions in a gas of primordial composition. We include a network of out of equilibrium reactions for the following ions and molecules: HI, HII, HeI, HeII, HeIII, H$_2$, H$_2^+$. The abundance of H$^-$ is approximated assuming the reactions are at equilibrium. The formation of H$_2$ includes the H$^-$ and the H$_2^+$ channels as well as 3-body reaction important at densities $>10^{10}$~\hcc. We include hydrogen and helium atomic cooling processes and an H$_2$ cooling function tested to be accurate up to densities $n_H \sim 10^{12}$~\hcc. We also model H$_2$ self-shielding using fitting functions from \cite{wolcottgreen2019} and the effect of secondary ionizations and heating from fast electrons produced by X-ray photoionizations \citep{shull1985, ricotti2002}. 
Since we do not use sink particles to create protostellar cores, in order to prevent artificial fragmentation at the maximum refinement level caused by a decreasing Jeans length, we suppress the cooling of cells with the maximum refinement level following the method in \citet{hosokawa2016}. This is done by multiplying the cooling function by the factor
\begin{equation}
    \label{eq:climit}
    C_{limit} = \exp{ \left[ -\left( \frac{\xi-1}{0.1} \right)^2 \right]} \hspace{0.5cm} \mbox{(if $\xi > 1$)},
\end{equation}
where $\xi = f_{limit} (\Delta x/\lambda_{J})$ with $\Delta x$ is the cell size and $\lambda_{J}$ is the Jeans length. We assume $f_{limit}=12$ as in \citet{hosokawa2016}. The masses of protostellar cores are therefore estimated by flagging cells with $C_{limit}<10^{-4}$.

X-ray self-shielding is neglected, but we run same tests showing that when including complete opacity to X-rays in cells with densities $\ge 10^{4}$~\hcc\ does not change any of the results we looked at. As \citet{hummel2015} has pointed out, however, the result depends on the total column density of the halo and the spectrum of the X-ray radiation. In future works more tests will be necessary to fully assess the importance of X-ray self-shielding in determining the masses and redshift of formation of Pop~III stars.
We also neglect HD cooling and chemistry. This could be important for strongly irradiated discs in which the H$_2$ fraction is large and the gas temperature reaches $T_{min}\sim 100-200$~K. HD cooling can reduce the characteristic mass of Pop~III stars \citep{yoshida2007} and its formation in a gas irradiated by X-rays so far has only been explored analytically \citet{nakauchi2014}.

Finally, in this paper we do not include radiative feedback from the accreating protostar, that is crucial in stopping the accretion onto the protostellar core and therefore in determining the final masses of Pop~III stars \citep{hosokawa2011,hosokawa2016,sugimura2020}. We estimate the final masses of Pop~III stars using 
an empirical relationship based on previous work by \citep{hirano2015} that includes UV radiation feedback. They provides a relation between the final mass in Pop~III stars, $M_{final}$, and the accretion rate onto the protostellar core, $dM/dt|_{cr}$, estimated at a characteristic radius:  
\begin{equation}
    M_{final} = 250 ~\mbox{M}_{\odot} \left( \frac{dM/dt|_{cr}}{2.8 \times 10^{-3} \mbox{M}_{\odot} \mbox{yr}^{-1}} \right)^{0.7}.
    \label{eq:finalmass}
\end{equation}
We define the characteristic radius when the collapsing core density reaches $n_H=10^7$\hcc\ where $M_{enc}(r)/M_{BE}(r)$ reaches its maximum value, consistently with the definition in \citep{hirano2015}. In Paper~I we find that the growth rate of the protostellar core is roughly constant as a function of time and is equal to $dM/dt|_{cr}$. 
Therefore we can also estimate the masses of Pop~III stars calculating their mass directly from the simulation at the characteristic time $\tau_{SF}$ defined as: 
\begin{equation}
\tau_{SF} = \frac{M_{final}}{dM/dt|_{cr}} = 89~{\rm kyrs}\left( \frac{dM/dt|_{cr}}{2.8 \times 10^{-3} \mbox{M}_{\odot} \mbox{yr}^{-1}} \right)^{-0.3}.
\end{equation}

One of the main results of Paper~I is that the accretion rate onto the protostar is  $dM/dt \sim dM/dt|_{cr} \propto c_s^3$, where $c_s$ is the minimum sound speed of the gas reached when the density is $n_H \sim 10^4$~\hcc. Therefore the timescale for star formation is $\tau_{sf} \propto c_s^{-1}$ and $M_{final} \propto c_s^2$. In terms of the minimum gas temperature we therefore have: $dM/dt \propto T_{min}^{3/2}$, $\tau_{SF} \propto T_{min}^{-1/2}$ and $M_{final} \propto T_{min}$.

\section{Results: I. Properties of Proto-stellar Discs}
\label{sec:frag}

Disc fragmentation is important in determining the number of Pop~III stars and their masses. Early cosmological simulations suggested that a single massive star ($\gtrsim 100$\msun) forms in each minihalo out of a pristine gas cloud \citep{abel2002, yoshida2008, hosokawa2011}, but more recent ones found the formation of multiple stars due to protostellar discs fragmentation \citep{clark2011,susa2014,stacy2016,sugimura2020}.
In our simulations, gas discs fragment to form multiple clumps once the central density reaches $10^{10} - 10^{11}$\hcc. In this section, we discuss the effect of background radiation on disc fragmentation.

\begin{figure}
    \centering
	\includegraphics[width=0.48\textwidth]{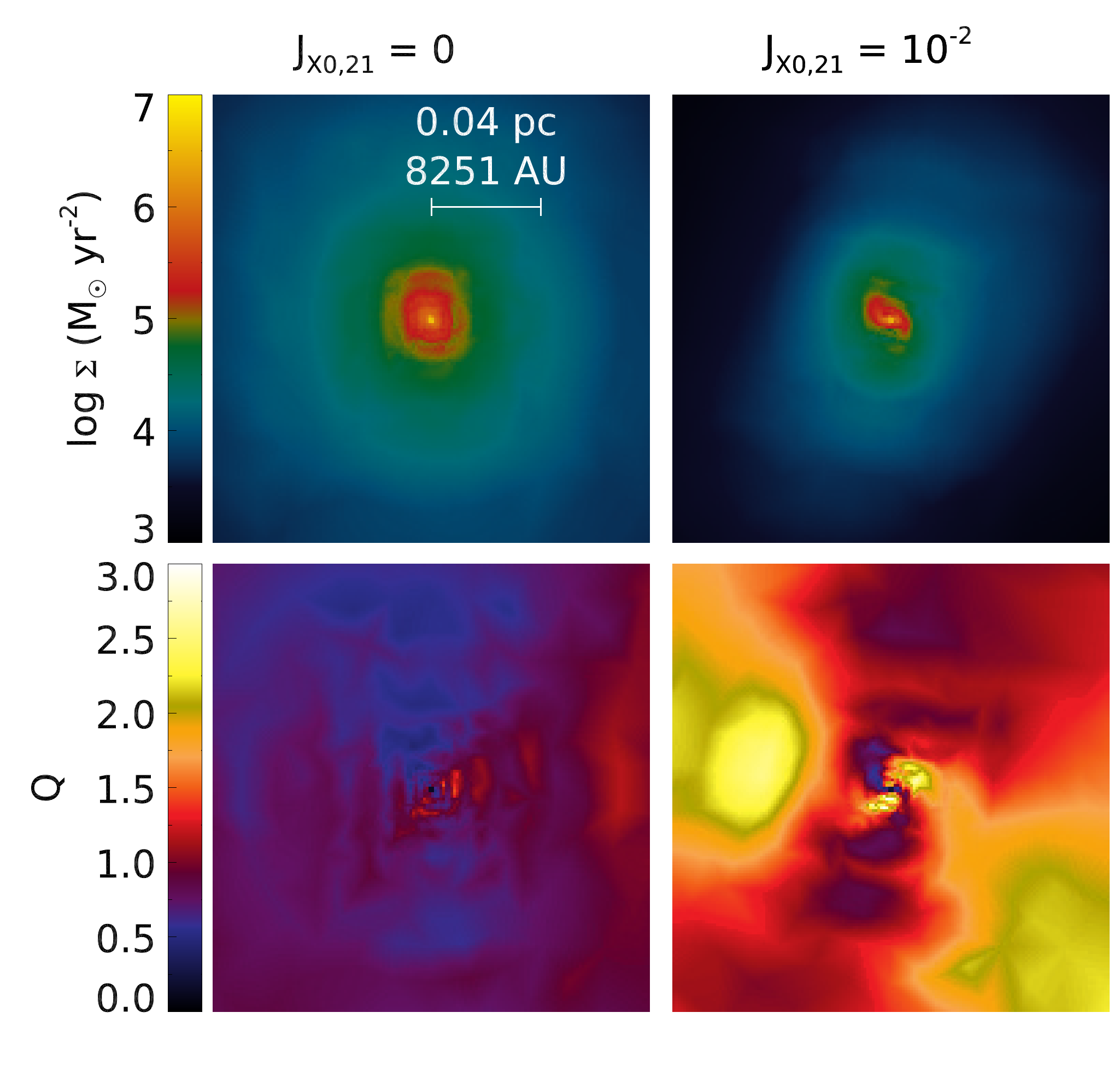}
    \caption{Surface density (top panels) and Toomre $Q$ parameter (bottom panels) of the face-on circumstellar disc in Halo~2, at the time when the first protostar forms at the centre. The left panels show the case without X-ray background and the right panels the case with a strong X-ray background with \jxz$=10^{-2}$. In both simulations \jlw$=0$, and the field-of-view is 0.16 pc $\times$ 0.16 pc (33000 au $\times$ 33000 au).}
    \label{fig:toomre_map}
\end{figure}

The main effect of X-ray irradiation on the properties of the disc is an increased stability to fragmentation. Fragmentation of a thin disc can be characterized by the Toomre $Q$ parameter defined as \citep{toomre1964}:
\begin{equation}
Q = \frac{c_s \kappa}{\pi G \Sigma},
\label{eq:Q}
\end{equation}
where $\kappa$ is the epicyclic frequency and $\Sigma$ is the disc surface density. If $Q>1$ the disc is stable to fragmentation. In Figure~\ref{fig:toomre_map} we show the comparison of the face-on view of two discs in Halo~2 without X-ray background (left column) and with an X-ray background of \jxz$=10^{-2}$ (right column). The top panels show the surface densities and the bottom panels the $Q$ parameter. It is clear that X-ray irradiated disc has lower surface density and larger $Q$ parameter.
\begin{figure}
    \centering
	\includegraphics[width=0.48\textwidth]{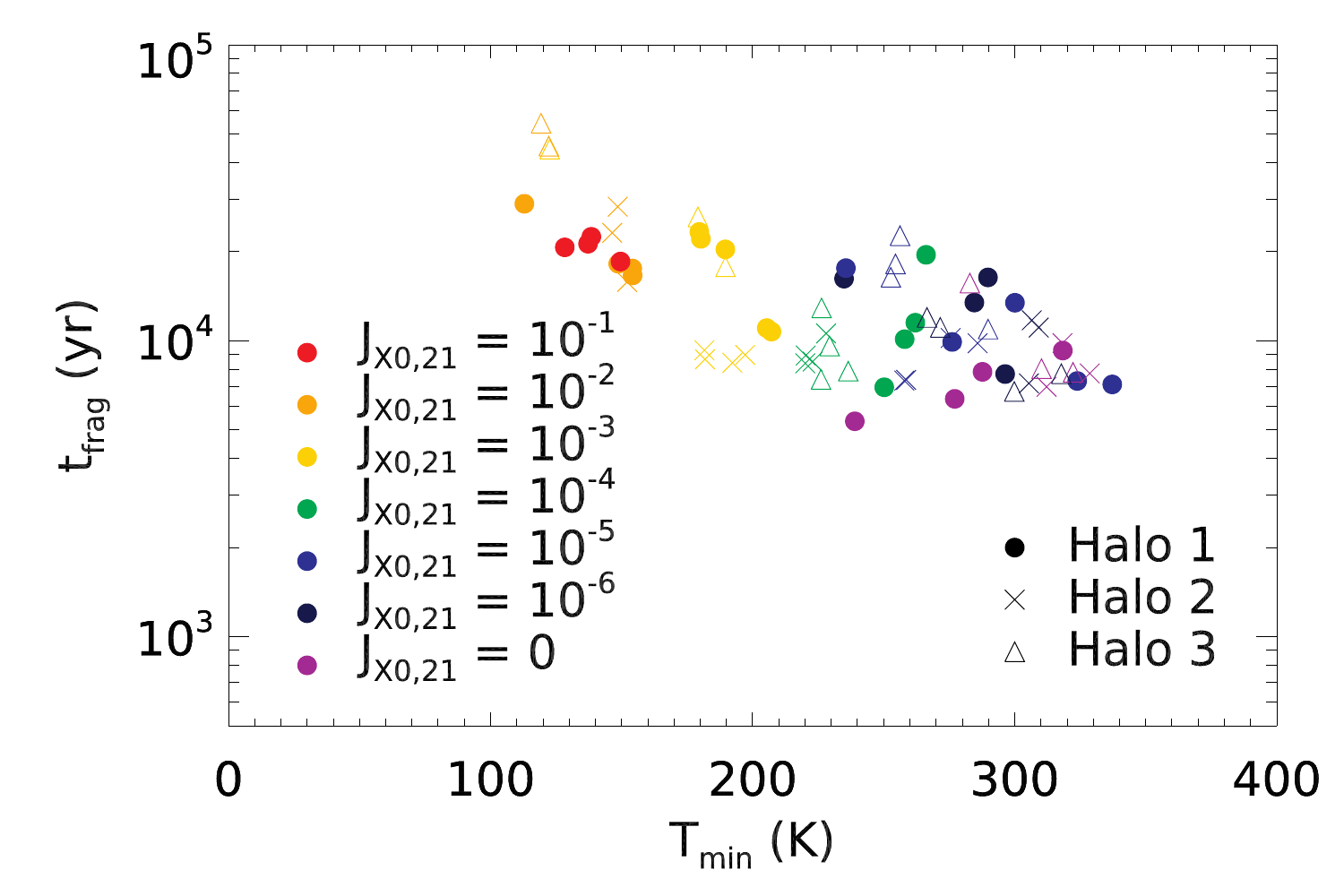}
    \caption{Fragmentation time as a function of the minimum gas temperature,  $T_{min}$, of the collapsing core in the gas phase diagram. \jxz\ is color-coded and the host halos are shown with different symbols (see legend). Simulations with halo critical mass $> 10^6$\msun\ are excluded from the analysis.}
    \label{fig:Tmin_tfrag}
\end{figure}
\begin{figure*}
    \centering
	\includegraphics[width=1.0\textwidth]{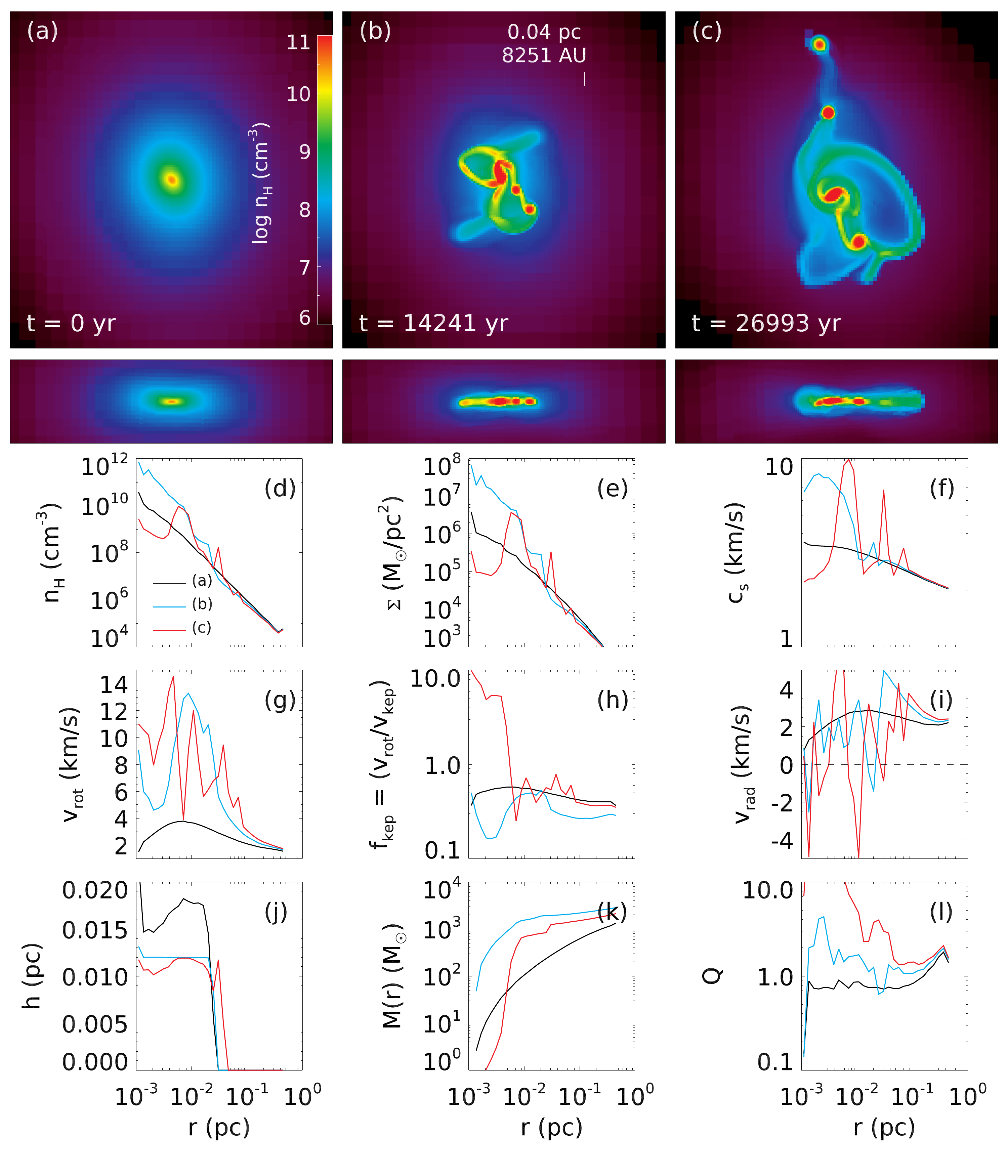}
    \caption{Disc of Halo~2 in the absence of a radiation backgrounds. Panel (a)-(c): Projected face-on (top) and edge-on (bottom) images of the disc in three snapshots at times $t=0, 14$ and 27~kyr, where $t$ is defined as the time since formation of the first protostar. The field-of-view of the face-on images is 0.16 pc $\times$ 0.16 pc (or 33000 au $\times$ 33000 au) and that of the edge-on images is 0.16 pc $\times$ 0.04 pc (or 33000 au $\times$ 8251 au). Panel (d): The radial profile of the hydrogen number density in the disc midplane. We plot the profiles of the three snapshots in panels (ab), (b) and (c) as solid lines with different colors as shown in the legend. Panel (e): The surface density of the gas. Panel (f): The sound speed. Panel (g): The rotational velocity. Panel (h): The ratio of the rotational velocity to the Keplerian velocity. Panel (i): The radial velocity. Panel (j): The thickness of the disc, defined as the value of $z$ where the gas becomes less dense than $5 \times 10^7$\hcc. Panel (k): The enclosed mass. Panel (l): The Toomre $Q$ parameter. }
    \label{fig:disk_zero}
\end{figure*}
\begin{figure*}
    \centering
	\includegraphics[width=1.0\textwidth]{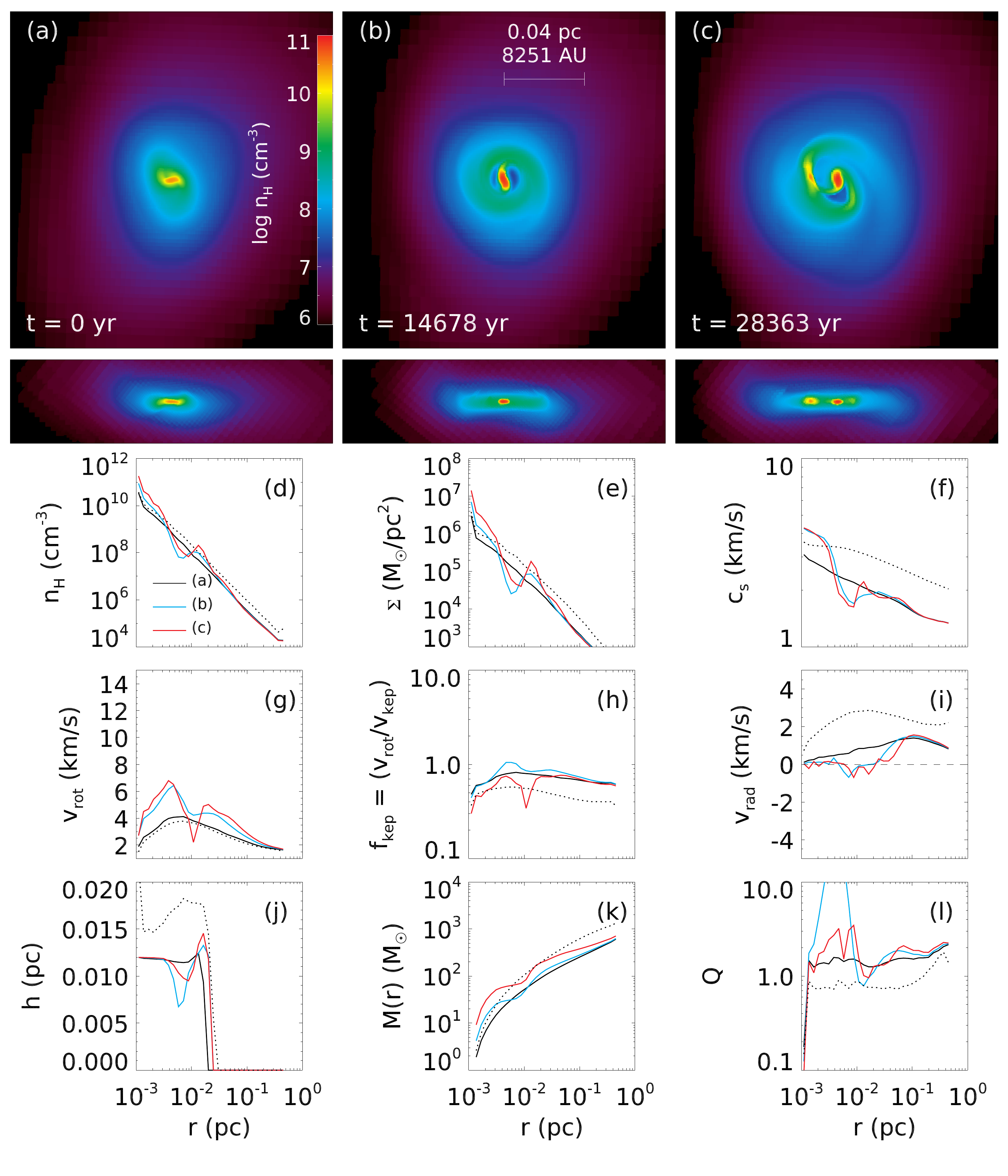}
    \caption{Same as Figure~\ref{fig:disk_zero} but for a disc irradiated by an X-ray background with intensity \jxz$=10^{-2}$ and \jlw=$0$. For the sake of comparison, the dotted black lines in Panels (d)-(l), show the case without X-ray irradiation at $t=0$.}
    \label{fig:disk_strong}
\end{figure*}

In addition to the parameter $Q$, another measure the stability of the disc is the fragmentation time $t_{frag}$, that here we define as the time between the formation of the first clump at the centre of the disc and the second clump. In Figure~\ref{fig:Tmin_tfrag} we plot its dependence on the minimum temperature, $T_{min}$, of the collapsing core in the phase diagram (see Paper~I). The plot shows that the second fragment forms systematically later in discs in a stronger X-ray background. This can be interpreted as due to a more stable disc (with larger $Q$), and/or the lower accretion rate $dM/dt$ found in a strong X-ray background.

Two examples of the evolution and fragmentation of discs with and without X-ray irradiation are shown in Figures~\ref{fig:disk_zero}-\ref{fig:disk_strong} for Halo~2. The top panels ((a)-(c)), show snapshots of the projected face-on and edge-on views of the disc at three different times. Panel (d) shows the hydrogen number density at the mid-plane of the disc as a function of $r$, while panels (e) to (l) show azimuthally averaged profiles in cylindrical coordinates of various quantities for the same three different snapshots (see legend). Figure~\ref{fig:disk_zero} shows the disc in the absence of any radiation background. When the first clump forms at the centre of the halo (panel (a)) the gas has a flat disc-like structure with central density $\sim 10^{10} - 10^{11}$\hcc. The disc starts to fragment $t \sim 14$ kyrs after the formation of the first clump (panel (b)), and at $t \sim 27$ kyrs four clumps can be identified (see panel(c)). As the disc fragments the azimuthally averaged profile of the disc shows fluctuations in the inner parts (cyan and red lines), hence the averaged profiles are less informative. Figure~\ref{fig:disk_strong} shows the same halo as Figure~\ref{fig:disk_zero}, but with an X-ray background of \jxz$=10^{-2}$. Fragmentation of the disc is suppressed for a longer time when compared to the case without X-ray irradiation. The disc starts to fragment after $t \sim 28$ kyrs (panel (c)) and the system remains a binary system for more than $30$~kyrs. In panels (d) to (l) we plot various properties of the disc and, for comparison, we show the case without X-rays at $t=0$ as dotted lines. One notable difference is that the surface density of the disc (panel (e)) and enclosed mass (panel (k)) become lower if the disc is irradiated by an X-ray background. The disc mass and surface density follow a similar trend as the mass of the collapsing hydrostatic core, that becomes lower in an X-ray background. Other notable difference is seen in the radial (infall) velocity (panel (i)). This follows from the lower accretion rate in a stronger X-ray background (and lower $T_{min}$).
Although the sound speed (panel (f)) and scale-height of the disc decrease with increasing X-ray irradiation, the disc is more stable to fragmentation due to the lower surface density, as shown by the Toomre $Q$ parameter (see panel (l)).

In order to interpret more quantitatively the properties of X-ray irradiated discs, below we derive simple scaling relationships for the main disc parameters as a function of X-ray irradiation. We aim at modelling the trends for the disc properties at early times, when the first protostar forms at the centre of the disc and before the disc fragments. These trends are shown in Figure~\ref{fig:disk_strong} as solid (strong X-ray) and dotted (weak X-ray) black lines ($t=0$, panel (a)).

We will model the disc properties in terms of the sound speed of the gas in the disc, that is related to the X-ray irradiation and $T_{min}$. As shown in panels (f) of Figures~\ref{fig:disk_zero}-\ref{fig:disk_strong}, the gas sound speed in the disc increases only by a factor of two with increasing density for densities between $10^4$~\hcc\ (at $r \sim 0.5$~pc) to $10^{10}$~\hcc\ (at $r \sim 10^{-3}$~pc). Therefore the gas is nearly isothermal with temperature directly proportional to $T_{min}$, defined as the minimum temperature reached by the collapsing gas at density $n_H \sim 10^4$~\hcc\ (see \S~\ref{sec:sim}). Note that panel (f) shows that, as the density approaches $10^{10}$~\hcc, the gas sound speed for the cases without X-ray and strong X-ray irradiation start to converge to a common value. This is in agreement with the results obtained in  former analytical studies \citep{Matsukoba2019,Kimura2021} showing that in the inner and denser regions of the disc the temperature reaches a thermal balance equilibrium between H$_2$ cooling and viscous heating. However, because viscous heating increases with increasing accretion rate, hence with decreasing X-ray intensity, the gas temperature in the inner parts of the disc also decreases with increasing X-ray irradiation, although the dependence is weak.

Assuming a thin disc in hydrostatic equilibrium with a central gravitational potential, and supported in the vertical direction by thermal pressure, the disc scale-height $H$ is:
\begin{equation}
\frac{H}{R} \sim \frac{c_{s}}{v_{kep}} \sim c_{s} (R/GM)^{1/2} \propto \mbox{const}(T_{min}),
\label{eq:scaleH}
\end{equation}
where $v_{kep}=(GM(<R)/R)^{1/2}$ is the Keplerian velocity, and we have assumed that the enclosed mass of the central clump $M(<R) \propto c_{s}^2$, scales as an isothermal sphere and $c_{s} \propto T_{min}^{1/2}$ as discussed above.
Hence, we expect that the disc scale-height is roughly independent of the X-ray intensity. Note that here we are only interested in the dependence of $H$ (and other quantities) on the X-ray irradiation and not the radial coordinate of the disc.

We expect that the midplane disc density roughly follows the isothermal cloud solution $n_H \propto c_s^2$, as suggested by $n_H \propto R^{-2}$ power-law of the density profile, that is very close to an isothermal sphere. The surface density profile 
\begin{equation}
\Sigma \propto n_{H}(R)H(R) \propto c_{s}^2 \propto T_{min}.
\end{equation}
also scales as $c_s^2$ since $H$ is roughly independent of $c_s$. In addition the gas accretion rate in our model (see Paper~I) is $dM/dt \propto \Sigma(R) v_{infall}(R) \propto c_s^3$, hence the gas infall rate should scale as $v_{infall}\propto c_s$.
The equations above imply that X-ray irradiated discs have lower surface density (and mid-plane density) and lower infall (radial) velocity than disc without X-ray irradiation. This indeed is what is observed in panels (d), (e) and (i) in Figure~\ref{fig:disk_strong} (see black solid and dotted lines).

Using the Toomre $Q$ parameter definition in Equation~(\ref{eq:Q}) we find:
\begin{equation}
Q \approx \frac{c_{s}^2}{\pi G \Sigma H} \frac{v_{rot}}{v_{kep}} \propto \frac{v_{rot}}{v_{kep}} \propto T_{min}^{-1/2},
\end{equation}
where we assumed $\kappa = \Omega = v_{rot}/R$ and that the rotational velocity of the disc is independent of the X-ray irradiation (as shown in panel (g) of Figure~\ref{fig:disk_strong}), and used equation~(\ref{eq:scaleH}) for the scale-height $H$. Therefore we expect that $Q$ increases and the disc becomes more stable (and with a more Keplerian rotation), with increasing X-ray irradiation (decreasing $T_{min}$).
Note that panel (h) in Figure~\ref{fig:disk_strong} supports the assertion that $f_{kep} \equiv v_{rot}/v_{kep}$, similarly to $Q$, increases with increasing X-ray irradiation.
Note that our approach differs slightly from previous modelling work of protostellar accretion discs around Pop~III stars. 
\cite{Matsukoba2019} relate the disc stability to the viscous $\alpha$ parameter in a quasi-Keplerian disc, assuming that $\alpha>1$ leads to large accretion rates and fragmentation. \cite{Kimura2021} argue that when the disc becomes more massive than the central star, the disc tends to experience fragmentation.  
Our model instead applies to early times ($t\sim 0$), when the disc may still be slightly contracting, before approaching a quasi-Keplerian rotation curve. However, our conclusions are qualitatively the same as we also find that discs with higher accretion rate (higher $T_{min}$), when the central protostar forms deviate more strongly from Keplerian rotation, and fragment more rapidly. X-ray irradiation leads instead to the rapid formation of a Keplerian accretion discs with lower accretion rate, and rather stable to fragmentation.
\begin{figure*}
    \centering
	\includegraphics[width=0.95\textwidth]{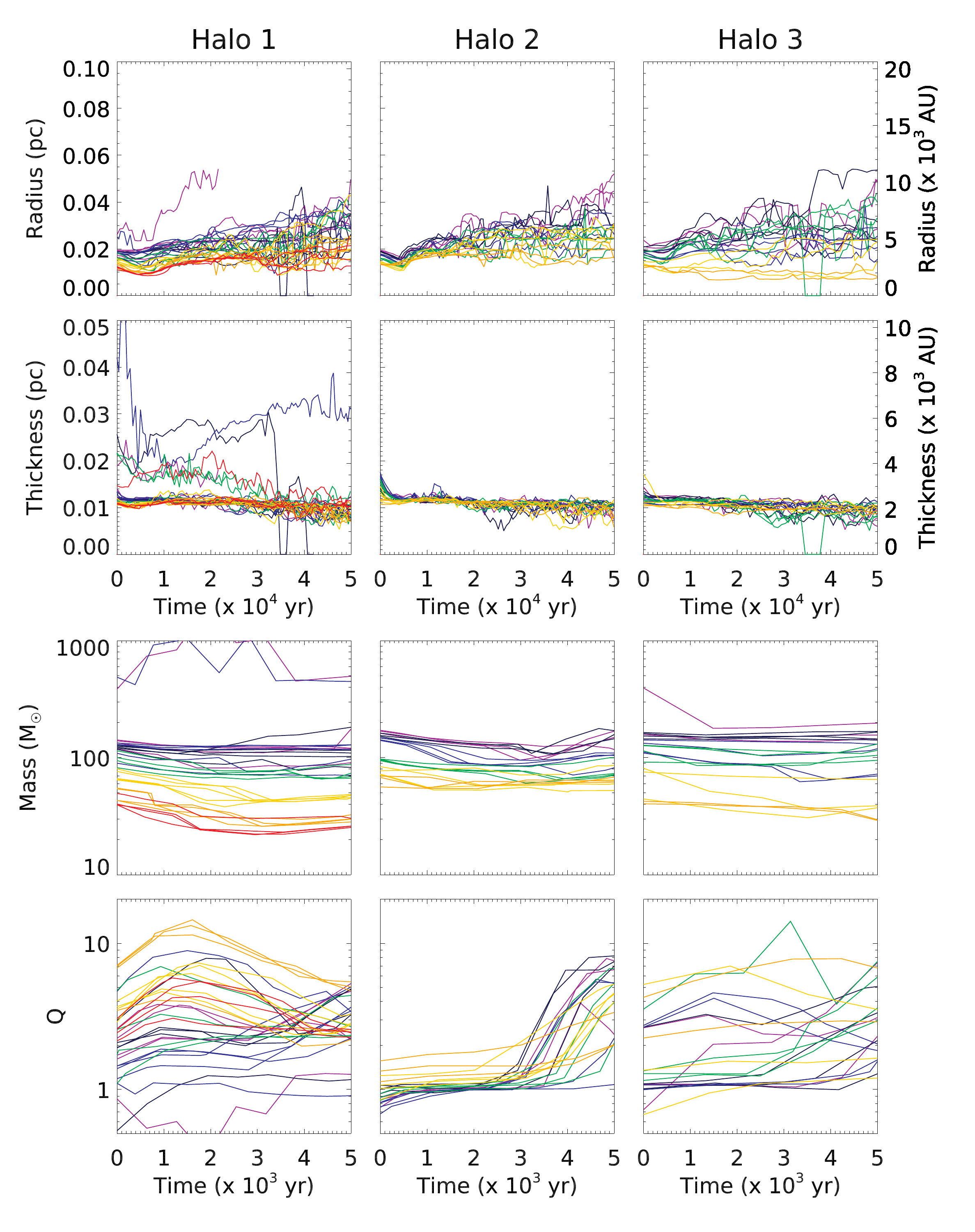}
    \caption{The radius, thickness, mass and average Toomre $Q$ parameter of discs in Halo~2 as a function of time. Each line show a simulation with X-ray intensity color-coded as in Figure~\ref{fig:Tmin_tfrag}. For the radius and thickness, we plot the results for 50~kyrs, while we only show the first 5~kyrs for the mass and $Q$ due to its large variations at later times due to disc fragmentation. The radius and thickness of the disc are defined as where the gas density drops below $5 \times 10^7$\hcc.}
    \label{fig:disk_evol}
\end{figure*}

Finally, we can estimated the disc mass as a function of X-ray irradiation. If we define the disc outer radius $R_{disc}$ where the surface density (or gas density) drops below a critical value, using a power-law fit with slope $1.5$ (panel (e) of Figure~\ref{fig:disk_zero} and \ref{fig:disk_strong}) for the surface density profile ($\Sigma(R_{disc}) \propto c_{s}^2/R_{disc}^{3/2}=\Sigma_{crit}$), we find:
\begin{equation}
R_{disc} \propto c_{s}^{4/3} \propto T_{min}^{2/3}.\label{eq:rdisc}
\end{equation}
Therefore, using the definition of disc radius above, the size of the disc decreases with increasing X-ray irradiation (decreasing $T_{min}$). The mass of the disc also decreases with increasing X-ray irradiation as:
\begin{equation}
\label{eq:mdisc}
M_{disc} \propto \Sigma(R_{disc}) R_{disc}^2 \propto R_{disc}^{2} \propto T_{min}^{4/3}.
\end{equation}
\hide{Since the mass in Pop~III stars roughly increases linearly with time, $t$: $M_{pop3} \sim dM/dt \times t \propto c_s^3 t$ and the final mass is $M_{final} \propto c_s^2 \propto T_{min}$ (see Paper~I), it is interesting to note that the mass ratio of the Pop~III stars to the disc mass scales as
$M_{pop3}(t)/M_{disc} \propto T_{min}^{1/6} t$ and $M_{final}/M_{disc} \propto T_{min}^{-1/3}$.Therefore with increasing X-ray irradiation we expect that at a fixed time, the disc mass is a larger fraction of the total mass in Pop~III stars, but given the longer duration of the accretion process in a strong X-ray background, the disc eventually becomes a smaller fraction of the final mass in Pop~III stars in a strong X-ray background \textcolor{red}{(KS: I don't understand why the disc mass can be assumed to be time independent here. While I think I now understand the argument on the disc structure at $t=0$ above, I'm still not sure about the analytical argument regarding the time evolution.)}.}

In order to check whether the simple scaling relationships derived above are consistent with the simulation results, in Figure~\ref{fig:disk_evol} in each row of panels, from top to bottom, we plot the disc radius, $R_{disc}$, the disc thickness, $h$, the disc mass, $M_{disc}$, and the average Toomre $Q$ parameter as a function of time. Different columns refer to the three halos and different \jxz\ are shown as lines of different color (see legend). 
The X-ray irradiation has an effect on the disc size, with smaller disc sizes in a stronger X-ray radiation background (the top row), in agreement with the analytic estimate in Equation~(\ref{eq:rdisc}).
The disc thickness, $h(R)$, shown in the second row, is defined the same way as in panel (j) of Figure~\ref{fig:disk_strong}: as the height above the disc midplane where the density drops below $5\times 10^7$~\hcc. The simulation results show that $h(R)$ is roughly constant as a function of disc radius $R$, while Equation~(\ref{eq:scaleH}) shows $H/R \sim {\rm const(R)}$. The disc thickness $h$ is related to $H$ as $h \propto H \times n_H(z=0)^{1/\beta}$, where we have approximated the density profile in the $z$-direction as a power-law: $n_H(z)\propto n_H(z=0)(1+z/H)^{-\beta}$. Assuming an isothermal profile for $n_H(z=0) \propto c_s^2/R^2$, and that $h(R)$ is constant as a function of $R$ as shown by the simulations, we find $\beta=2$ and $h(R) \propto (H/R) c_s \propto c_s$. This results is in qualitative agreement with panel (j) of Figure~\ref{fig:disk_strong} and with the early time evolution ($t\sim 0$) of the disc thickness shown in the second-row panels in Figure~\ref{fig:disk_evol}. However, at later times the mean disc thickness $h$ becomes nearly independent of the X-ray intensity and remains rather constant as a function of time for 50~kyrs.
The third row of panels in Figure~\ref{fig:disk_evol} show the discs masses become smaller with increasing X-ray irradiation, as expected by Equation~(\ref{eq:mdisc}). In the bottom panels we plot the average $Q$ parameter as a function of time for the first 5~kyrs of the disc evolution. As from the analytic model the figure shows that the disc is smaller, has lower surface density and is more stable (larger Toomre $Q$ parameter) with increasing intensity of the X-ray background. Especially in Halo~2 (and to a lesser extent for Halo~1), that experiences slow growth during star formation, this trend is clear: in a weak X-ray background $Q<1$ and increases as the disc is more strongly irradiated by X-rays. In Halo~3 the trend is less clear, probably due to its rapid growth. As \citet{hummel2015} points out, the growth rate of the minihalo is also an important factor for the fragmentation of the protostellar disc, and can wash out the dependence on the X-ray intensity.
\begin{figure*}
    \centering
	\includegraphics[width=0.95\textwidth]{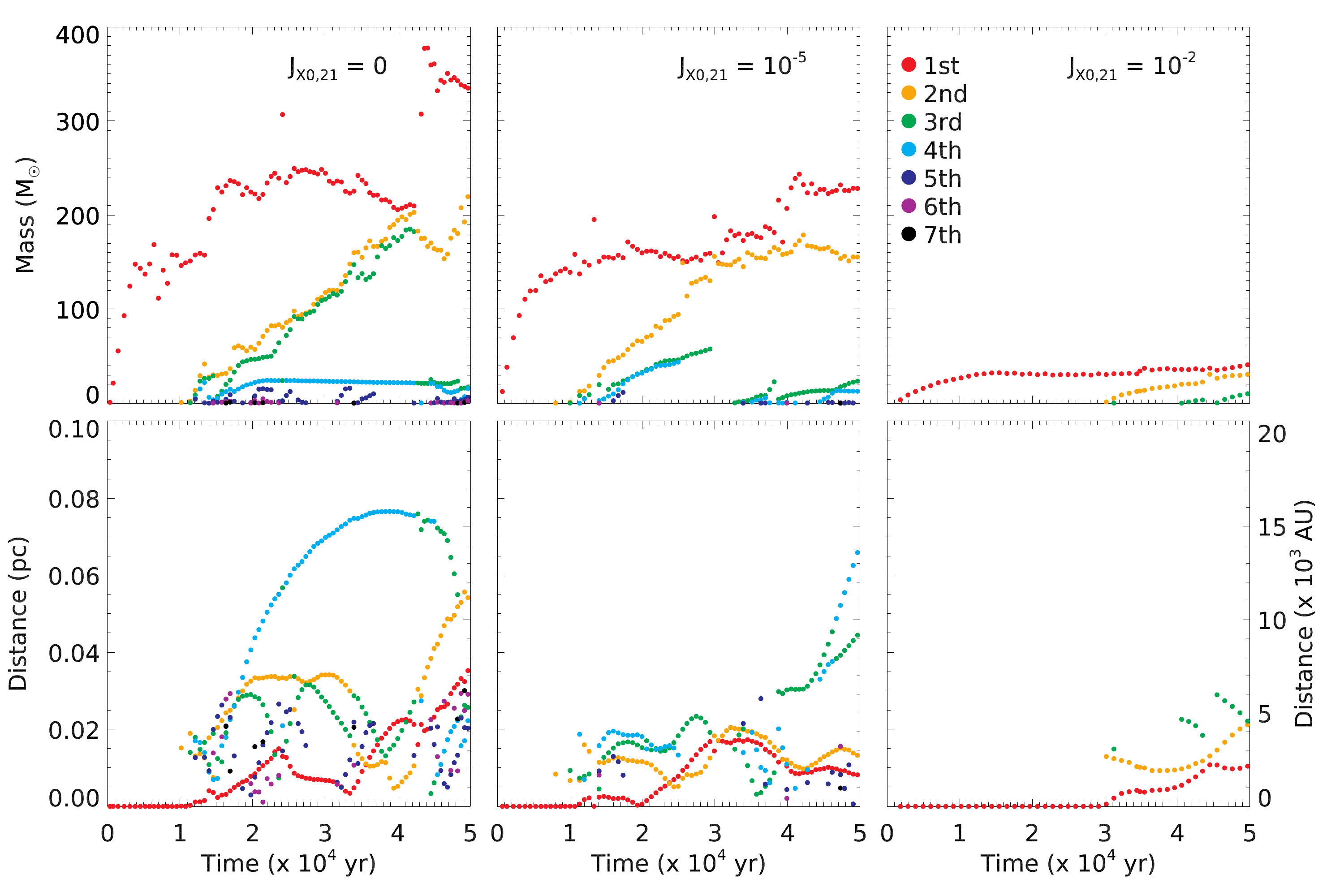}
    \caption{Masses (top panels) and distances from the centre (bottom panels) of individual clumps in Halo~2 as a function of time after the formation of the first clump. From left to right, \jxz\ is $0, 10^{-5}$ and $10^{-2}$\jj. The masses and positions of protostars (disc fragments) are color-coded as shown in the legend, from the most massive to the least massive fragment.}
    \label{fig:evol_ind}
\end{figure*}
\begin{figure}
    \centering
	\includegraphics[width=0.45\textwidth]{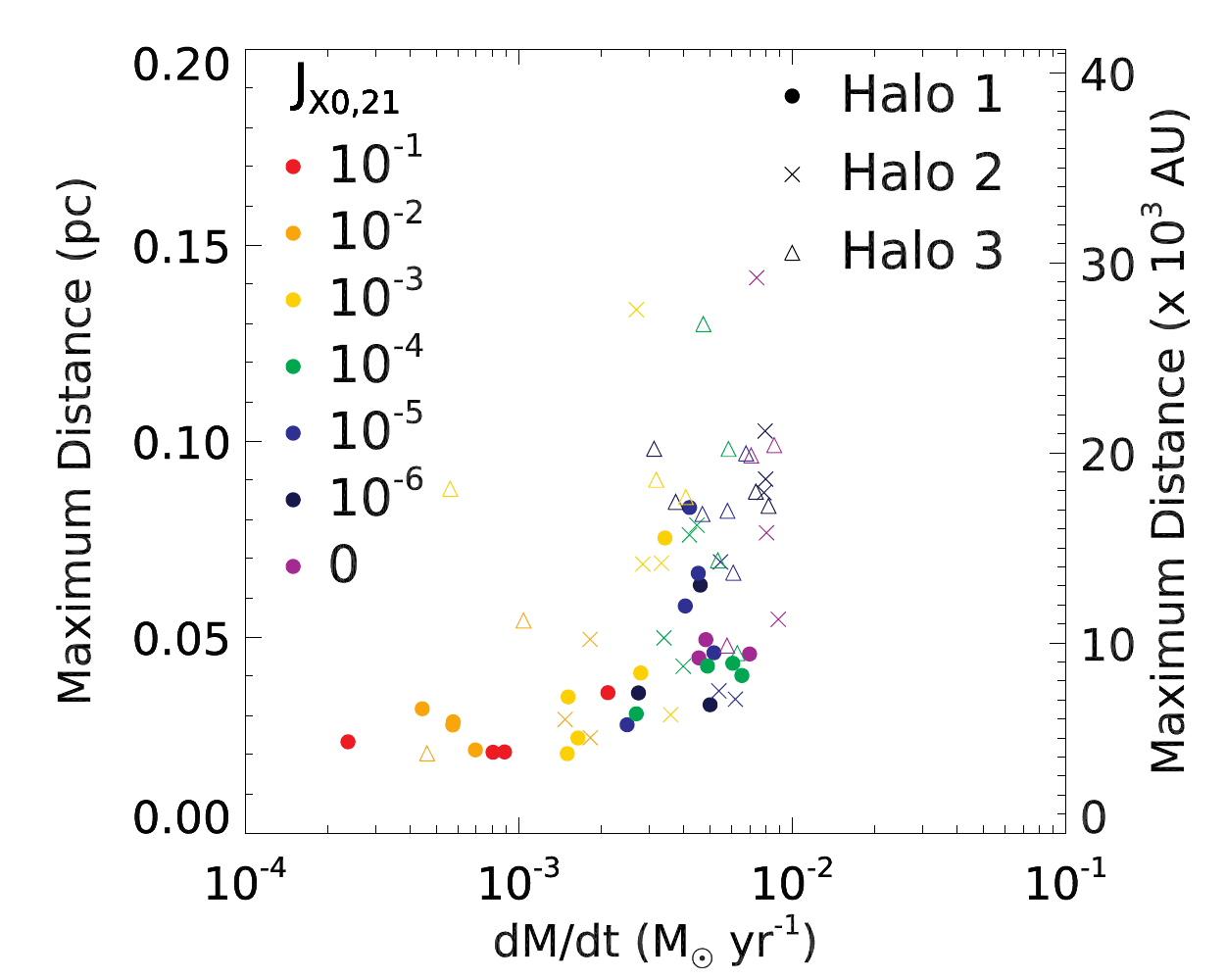}
    \caption{Maximum distance of a clump as a function of the gas accretion rate at $\nh=10^7$\hcc, $dM/dt|_{cr}$. The symbols and color of each point indicate the intensity of the X-ray radiation background and host halo, as in the legend. Only simulations with the critical mass lower than $10^6$\msun\ are shown in the plot.}
    \label{fig:Mdot_dist}
\end{figure}
\begin{figure}
    \centering
	\includegraphics[width=0.48\textwidth]{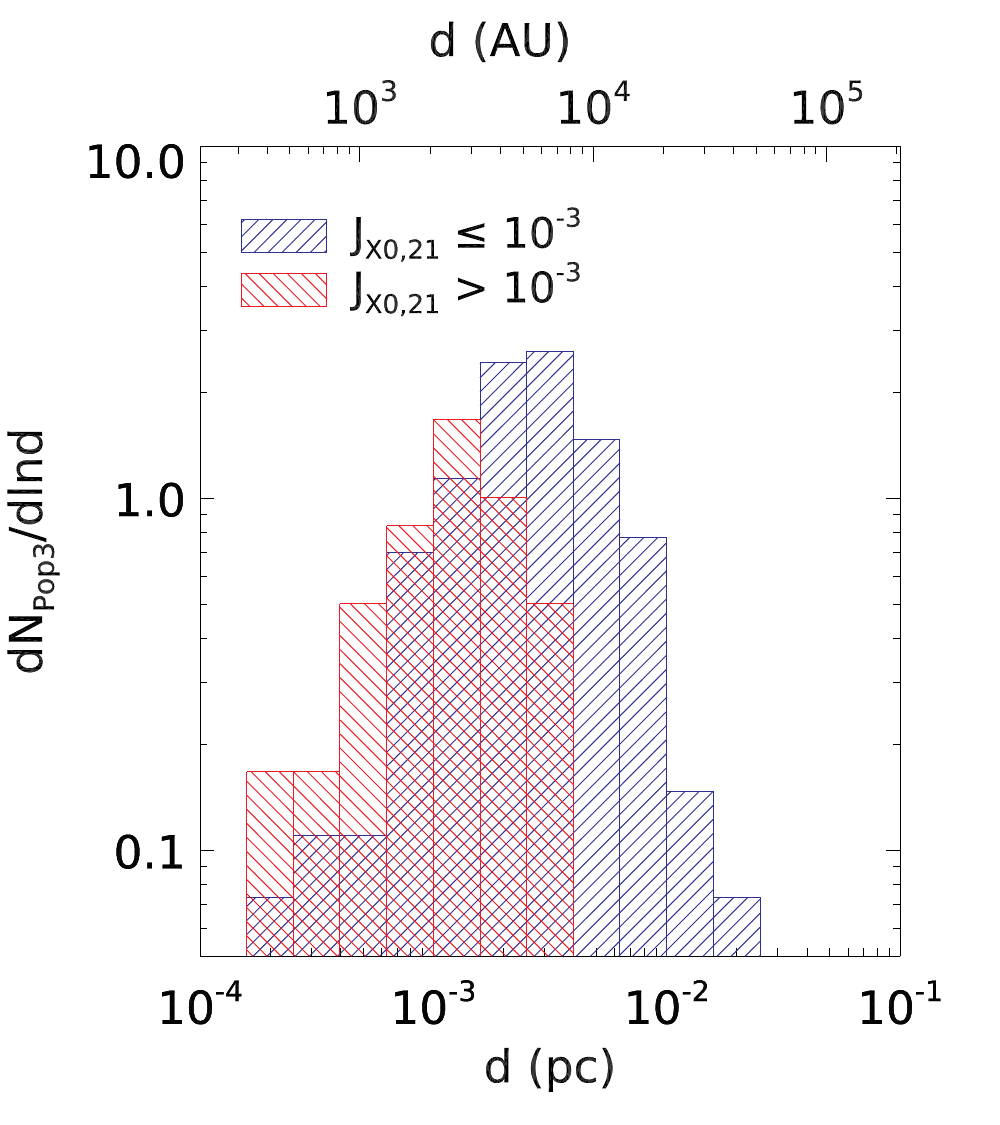}
    \caption{Histogram of the distances of Pop~III stars from the centre-of-mass in a weak (\jxz$\leq10^{-3}$, blue) and strong (\jxz$> 10^{-3}$, red) X-ray background. The distances are measured either when the total mass of the stars is equal to $M_{final}$ (see Equation~\ref{eq:finalmass}) or at $t=5\times10^4$~yrs if the total mass does not reach this limit. Pop~III stars forming in host halos more massive than $10^6$\msun\ are not included in this plot.}
    \label{fig:dist_hist}
\end{figure}
\section{Results: II. Multiplicity and Migration of Population~III stars}\label{sec:mult}

Figure \ref{fig:evol_ind} shows the masses (top panels) and the distances from the disc centre (bottom panels) of individual clumps in Halo~2 for different X-ray backgrounds. The mass of the most massive clump (red dots) tends to be smaller in a stronger X-ray background, similarly to the total mass. Disc fragmentation occurs $\sim 10$~kyrs after the formation of the first clump for the cases with no or weak X-ray background (left and middle panel), while it occurs at $\sim 30$ kyrs in a strong X-ray background (right panel).
\begin{figure*}
    \centering
	\includegraphics[width=0.9\textwidth]{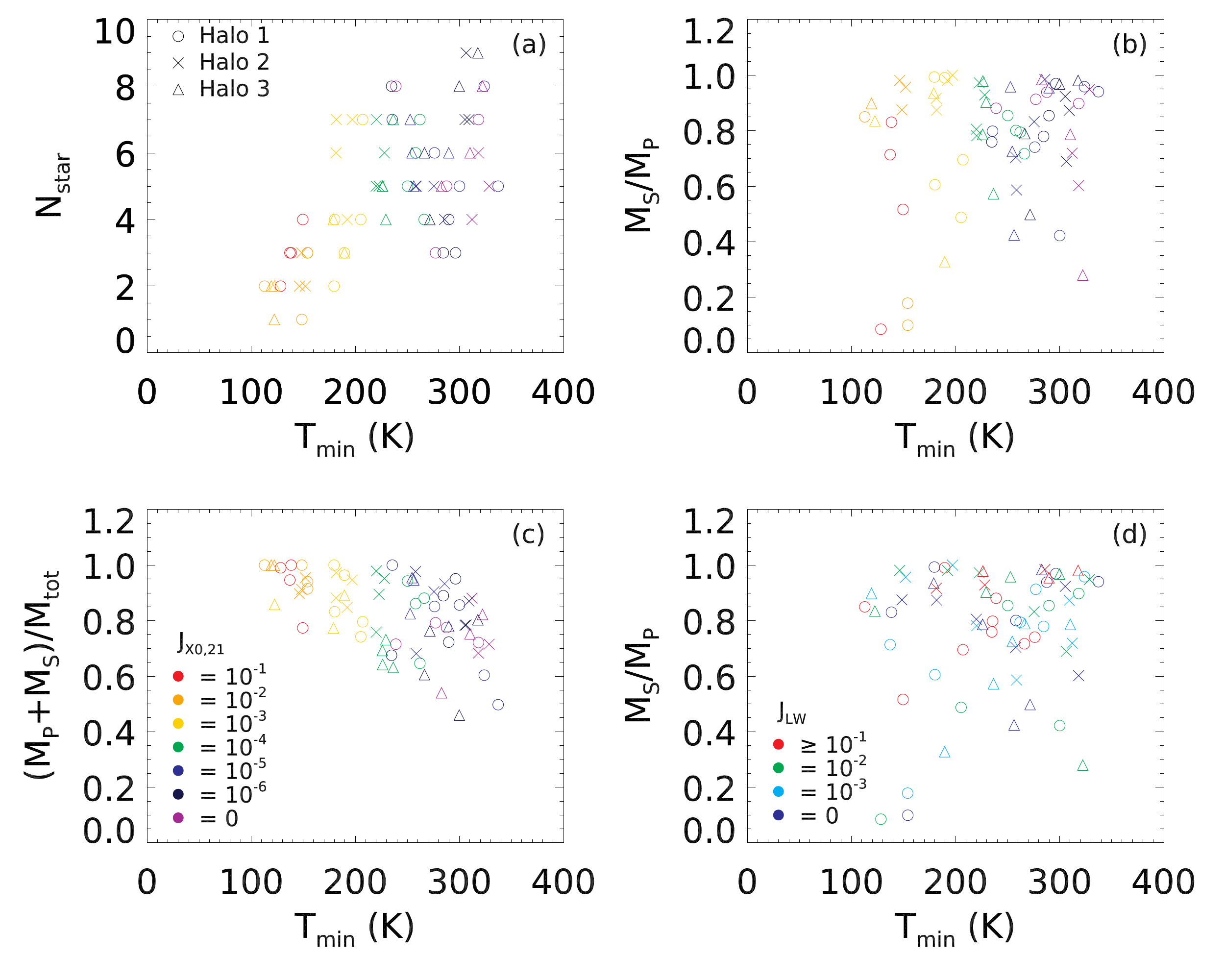}
    \caption{Panel (a): Total number of clumps when the total mass exceeds $M_{final}$, including clumps that merged with others. The symbol and color of each point indicates the intensity of the X-ray background, \jxz, and the host halo, as shown in the legend. Results with the critical masses $> 10^6$\msun\ are omitted. Panel (b): The mass ratio of the two most massive (primary, $M_{P}$, and secondary, $M_{S}$) stars. Panel (c): the ratio of $M_{P}+M_{S}$ to the total mass. Panel (d): Same as Panel (b) but the colors indicates the intensity of the LW background.}
    \label{fig:frag}
\end{figure*}
\begin{figure*}
    \centering
	\includegraphics[width=0.48\textwidth]{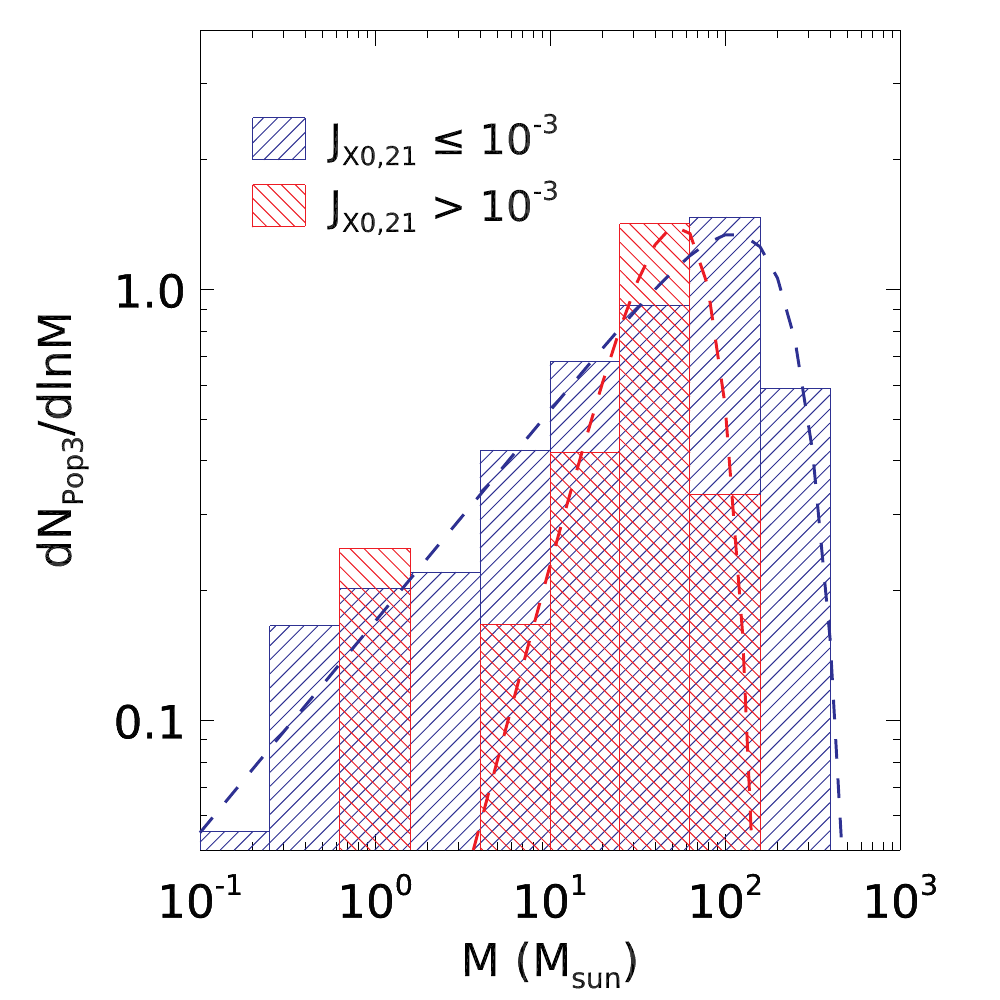}
	\includegraphics[width=0.48\textwidth]{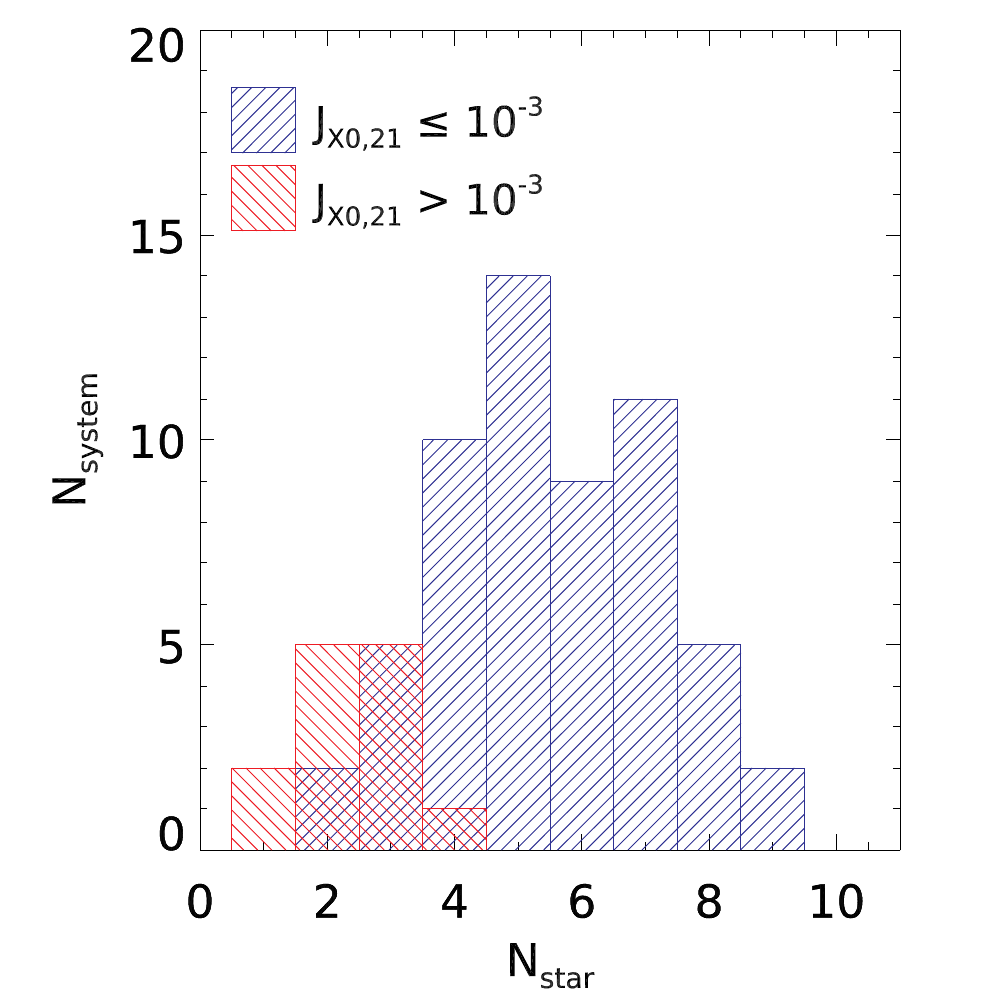}
    \caption{Mean mass function per halo (left panel) and distribution of the multiplicity of stars in each halo, $N_{star}$ (right panel) in a weak (blue histograms) and strong (red histogram) X-ray background. The dashed lines in the left panel show fits to the mass functions with Equation~(\ref{eq:IMF}). The parameters of the fit are shown in Table~\ref{tab:IMF}, including 
    the total mass in Pop~III stars, $M_{tot}$, and the average multiplicity per halo $N_{tot} = \int_{\ln 0.1\,M_\odot}^{\ln 1000\,M_\odot} (\mathrm{d} N_{pop3}/\mathrm{d\ln M})\,\mathrm{d}\ln M$. The values $N_{tot}\sim 2$ and $\sim 4$ found for the strong and weak X-ray background, respectively, are consistent with the multiplicity distribution shown in the right panel.}
    \label{fig:imf}
\end{figure*}
\begin{figure}
    \centering
	\includegraphics[width=0.48\textwidth]{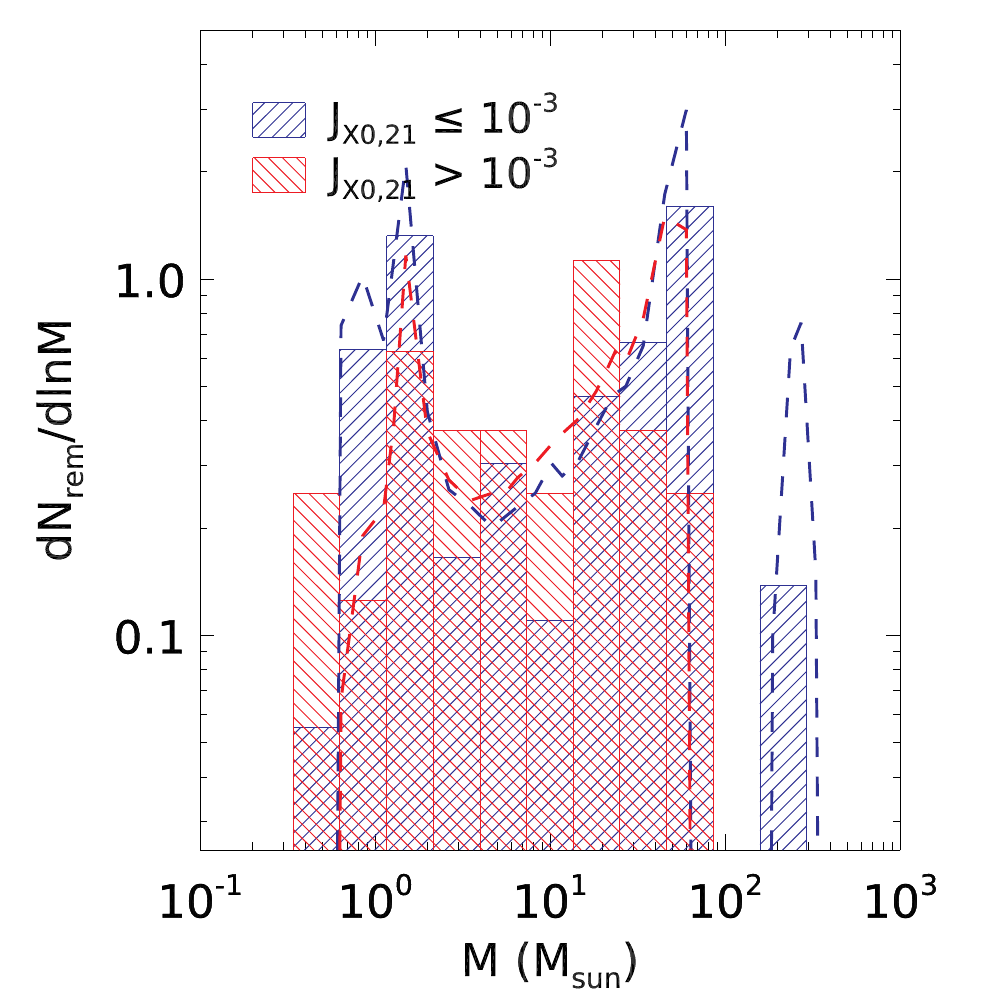}
    \caption{The mass functions of compact remnants (NS and BHs) in the two samples corresponding to the mass functions in Figure~\ref{fig:imf}. We converted the masses of stars into those of the remnants using  Equation~(\ref{eq:remnant}) from results in \citet{heger2002}. The dashed lines show the remnant's mass functions obtained using Monte-Carlo simulations with $10^5$ stars. The mass functions are normalized consistently with Figure~\ref{fig:imf}, so that the total initial mass in Pop~III stars is equal to the mean mass in Pop~III star per halo in each group.}
    \label{fig:remmf}
\end{figure}
One important aspect of the formation of Pop~III star is the survival of the stars after disc fragmentation. 
Previous simulations showed that while the inward migration of secondary stars leads to the merger with the primary stars, protostars are also observed to migrate outward \citep[see][]{greif2012,stacy2013,stacy2016,hirano2017,Susa2019,Chon2019,sugimura2020}.
The star migration is affected by gas in the disc through two main processes: i) accretion of gas with higher angular momentum of the protostar produces an outward migration (in a Keplerian disc it's the gas with orbit outward of the protostar orbit); ii) dynamical friction and gravitational torques mediated by resonances, 
produce a migration inward of the protostars by
loss of their angular momentum and energy to the disc. Typically the first process is dominant when the disc is thick, such as the zero-metallicity discs in this study \citep{HeatN:2020, sugimura2020}. In bottom panels of Figure~\ref{fig:evol_ind} we plot the distances of individual clumps from the centre-of-mass. In each panel, the most massive clump (red dots) forms first, at the centre of the disc so its initial distance is zero. As the disc fragments and the second clump forms and grows, the primary and secondary protostars both migrate outward from the centre. In zero or weak X-ray backgrounds (left and middle panel), several clumps migrate beyond $10^4$~AU while no one migrate to such large distance for $\sim 80$~kyr in a strong X-ray background. We speculate this is related to the property of the protostellar disc. The gas accretion of high angular momentum gas into the protostar is slow and therefore clumps do not migrate rapidly outward. In addition there are fewer stars that cause outward migration through angular momentum transfer \citep[see][]{greif2012,stacy2013}. We plot the maximum distance of clumps in each system as a function of the accretion rate at $\nh=10^7$\hcc\ in Figure~\ref{fig:Mdot_dist}. The large vertical scatter of symbols for $dM/dt|_{cr} \sim 2\times10^{-3}-10^{-2}$, indicates migrations beyond $10^4$~AU are common in a weak X-ray background. On the contrary, outward migration is rare in a strong X-ray background and the maximum distances are smaller than $10^4$~AU. Figure~\ref{fig:dist_hist} shows the probability distribution of distances from the disc centre of protostars in a weak (\jxz$\leq10^{-3}$, blue shaded histogram) and strong (\jxz$>10^{-3}$, red shaded histogram) X-ray background. The distances are measured either when the total mass of the stars is equal to $M_{final}$ (equation~(\ref{eq:finalmass})) or at the end of the simulation at $t=5\times10^4$~yrs if the total mass does not reach this limit. Stars in host haloes more massive than $10^6$\msun\ are not taken into account in this analysis. The two distributions are rather similar to each other and symmetric, but the peaks are offset by $\sim 3000$~AU: at $\sim 6000$~AU and at $\sim 3000$~AU in weak and strong X-ray background, respectively. This result also suggests that accretion rate onto the protostar drives the outward migration. 

To study the IMF we need to determine the masses of individual stars. Unlike the total mass, however, individual clumps do not always grow linearly. For this reason we cannot use the same approach we use in the Paper~I to estimate the total mass. Instead define the clump masses at the time when the total mass reaches $M_{final}$ or, if this condition is not met, at the end of the simulations: $\sim 50$~kyrs after the formation of the first clump. We define $M_{P}$ as the the mass of most massive star (the primary star) and $M_{S}$ as the mass of the second most massive star (or the secondary star). We also count the total number of clumps that form in the discs. If clumps form and merge to form more massive ones within three snapshots interval ($\sim 2-3$~kyr), we do not count them as individual clumps. Although this criterion is rather arbitrary, the number and merger history of small mass clumps and/or close-binaries is intrinsically uncertain, as it also strongly depend on the initial conditions and the resolution of the simulations.

Panel (a) of Figure~\ref{fig:frag} shows that the number of clumps decreases with increasing X-ray intensity. The gas disc is more stable in a stronger X-ray background and therefore undergoes fewer fragmentations. In the absence of both X-ray and LW radiation backgrounds, for instance, six clumps form out of the protostellar disc in Halo~2  (one of the clumps merge with a more massive one shortly after its formation). When the total mass reaches $M_{final}$ ($\sim 33$~kyrs after the first clump formation), the system is composed of three massive clumps ($M > 100$\msun) and two small ones ($M < 20$\msun). On the other hand, in a strong X-ray background, a binary system is formed at $\sim 50$~kyrs. Two additional smaller mass clumps form in this simulation, but they quickly merge with the other two. 

Panel (c) shows the ratio of $M_{P}+M_{S}$ to the total mass. As \jxz\ increases, the mass ratio converges to 1. This is consistent with the trend of N$_{star}$ in panel (a), because fewer stars form in a strong X-ray background, and thus the primary plus secondary stars account for most of the mass. The range of the mass ratio gets wider in a weak X-ray background because of the large scatter in the number of stars.

Hence, the multiplicity of stars is related to the intensity of the X-ray background, but what is the dependence of the mass function of the clumps on the X-ray background? Panel (b) of Figure \ref{fig:frag} shows $M_{S}/M_{P}$ as a function of $T_{min}$. In most simulations the primary and secondary stars have similar masses ($M_{S}/M_{P}\sim 0.8-1$), but the scatter is large and, considering all the haloes, no clear trend with X-ray intensity is found. However, if we exclude Halo~1 from the analysis (circles), there is a preference for forming equal mass binary stars in a strong X-ray background. We have seen before that some trends in Halo~1 are less clear due to its rapid accretion rate and a merger with a satellite clump during Pop~III formation. To investigate the role of LW background on the mass function, we also color-code the symbols according to the LW intensity in Panel (d). With low LW intensity (\jlw$\leq 10^{-2}$), symbols are widely distributed and show no significant correlation between the backgrounds and the mass ratio. On the other hand, when the LW background is stronger than or equal to $10^{-1}$, the two most massive stars tend to be of similar mass.

The trends found so far are more quite clear if we bin all the simulations results for the three haloes in two groups: haloes in a strong and weak X-ray background.
In the left panel of Figure~\ref{fig:imf} we show the mass function of the fragments (protostars) in weak (\jxz$\leq10^{-3}$, blue shaded histogram) and strong (\jxz$>10^{-3}$, red shaded histogram) X-ray background.

Both mass functions are top-heavy and they can be described as a power-law with an exponential cutoff,
\begin{equation}
    \frac{dN}{d\ln M} = A M^\alpha \exp{\left[-\left(\frac{M}{M_{cut}}\right)^2\right]}.
    \label{eq:IMF}
\end{equation}
Parameters of each mass function are shown in Table~\ref{tab:IMF} with the peak mass ($M_{peak}=M_{cut} \sqrt{\alpha/2}$). They are normalized so that the total mass in each distribution is the mean mass per halo and the the total number of stars is the mean multiplicity (i.e., dividing the total mass by the number of runs in each group). The mass function in a strong X-ray background is steeper (higher $\alpha$) and has lower peak mass. These differences can be explained by the following two factors. First, the protostellar disc experiences fewer fragmentations and therefore there are fewer low-mass stars in a strong X-ray background. Secondly, in a strong X-ray background, the lower accretion rate leads to a smaller total mass disc stars, $M_{final}$, and a smaller primary star mass ($M_{P}$). Hence both the cutoff mass an the total mass in fragments is lower by about a factor of $2-3$ in a strong X-ray background.
\begin{table}
	\centering
	\caption{Parameters of Mass Functions}
	\begin{tabular}{ | c | c | c | c | c | c | c |}
		\hline
		\jxz & $\alpha$ & $M_{cut}$ & $A$ & $M_{peak}$ & M$_{tot}$ & N$_{tot}$\\
		\hline
		$\leq 10^{-3}$ & 0.490 & 229\msun & 0.169 & 113\msun & 341\msun & 4.39 \\
		\hline
		$> 10^{-3}$ & 1.53 & 61\msun & 0.00692 & 53.4\msun & 102\msun & 2.25 \\
        \hline
	\end{tabular}
	\label{tab:IMF}
\end{table}
Note that these are mass functions created from multiple stellar systems in only three minihaloes, but for a large number of simulations with different combinations of the X-ray and LW backgrounds. Hence the variance on the total mass and the mass of the most massive fragment may be not representative of all Pop~III stars. Also, we have used a simplistic prescription to take into account the effect of feedback in determining the final masses of the fragments, that may not be accurate. Nevertheless, the distribution should be a rather accurate representation of the initial mass function of the disc fragments, before longer timescale dynamical processes destroy the systems be either ejecting stars or merging them or forming tight binary systems.

The right panel in Figure~\ref{fig:imf} shows the distribution, ${dN_{system}}/{d \ln N_{star}}$, of the stellar multiplicity, $N_{star}$, of each system (minihaloes). All the systems in a strong X-ray background are either binary or triple/quadruple systems. On the contrary, in a weak X-ray background most of the systems have multiplicity greater than five.

\section{Summary}
\label{sec:disc}

It has been recognized for several years that the first metal-free stars forming in the universe have a top-heavy IMF \citep{bromm2001, abel2002, bromm2002}.
Hence, the first stars can produce powerful hypernova explosions detectable with James Webb Space Telescope and Nancy Grace Roman Space Telescope \citep{whalen2014}, produce a population of IMBHs that may contribute to the gravitational wave detections in LIGO \citep{abbott2016}. In addition strong or weak feedabck from Pop~III star explosions and metal enrichment they produce, may have important effects on the formation of Pop~II stars in the first galaxies \citep{RicottiGS:2002b,RicottiGS:2008, wise2008, greif2010, Abe2021}.

LW and X-ray photons emitted by Pop~III stars and their remnants can travel large distances without being absorbed and build a radiation background. Ionization by the X-ray background produce high-energy photo-electrons and in a mostly neutral gas a large fraction of their kinetic energy is deposited into secondary ionization \citep{shull1985}. These backgrounds regulates \htwo\ formation on cosmological scales and thus the formation of early objects \citep[first stars and their remnants such as SNe and HMXBs,][]{venkatesan2001,jeon2014,ricotti2016,xu2016}. The effects of a LW radiation background on Pop~III star formation have been studies rather extensively \citep{haiman2000,omukai2001,hirano2015,regan2020}, instead the role of the X-ray radiation background has received less attention and the impact X-rays have on the formation rate and the IMF of the first stars is still debated \citep{jeon2014,hummel2015,ricotti2016}.

In this study, the second in a series, using zoom-in simulations of three minihaloes with different mass and irradiated by different intensities of the LW and X-ray backgrounds, we investigate the effect radiation backgrounds on the properties of protostellar discs and their fragmentation, therefore deriving the multiplicity and mass function of the first stars. Below we summarize the key results of the simulations.
\begin{enumerate}
    \item X-ray radiation affects the properties of protostellar discs and their fragmentation. The discs have lower surface density, they are smaller in size and mass and are more stable to gravitational instabilities (i.e., they have larger Toomre-$Q$ parameter).
    \item Pop~III stars in a weak or absent X-ray background form multiple systems of $6 \pm 3$ stars, with masses around $100$\msun\ and typical distance from the centre of mass of 2,500-20,000~au.
    \item Pop~III stars in a strong X-ray background typically form binary systems of nearly equal mass stars with individual masses around $40$\msun\ and distances from the centre of mass of 1,000-5,000~au.
    \item Independently of the strength of the X-ray background, protostars form near the centre of the disc (within 5000~au) and migrate outward while growing in mass. The migration is mainly driven by accretion of gas with higher specific angular momentum from the outer parts of the nearly Keplerian disc. 
    \item We find that the mass function of the fragments is well described by a power law with positive slope and an exponential cutoff. In a strong X-ray background the mass function of protostars is slightly more to top-heavy, with slope $\alpha=1.53$ and mass cutoff of 61~\msun. While for weak or no X-ray background $\alpha=0.490$ and $M_{cut}=229$~\msun.
\end{enumerate}
The final fate of Pop~III stars of a given mass is still rather uncertain, but following the work of \citep{heger2002,nomoto2006} we will assume that Pop~III stars with masses in the range $40~M_\odot<M<140$\msun\ and $>260$\msun\ collapse directly into BHs without releasing mechanical energy and metals by SN explosions. Instead stars with masses $140~\mbox{M}_{\odot}<M<260$\msun\ explode as PISNe without leaving any remnant. Finally stars with masses $11~\mbox{M}_{\odot}<M<20$\msun\ explode as normal SNe and stars with masses $20~\mbox{M}_{\odot}<M<40$\msun\ as hypernovae.

Given the caveats on the mass function discussed in Section~\ref{sec:mult}, and the uncertainties on the final fate of Pop~III stars of a given mass, it is nevertheless interesting to estimate the energy in SN explosions and the mass of BH remnants for the strong and weak X-ray irradiation cases.
Following \citet{wise2012} we will assume that 
\begin{equation}
\frac{E_{SN}(M)}{10^{51} \mbox{ erg}}=
\left\{
	\begin{array}{ll}
		1  &  {\rm if}~11~\mbox{M}_{\odot} \leq M < 20~\mbox{M}_{\odot}\\
		-13.714+1.806 M & {\rm if}~20~\mbox{M}_{\odot} \leq M < 40~\mbox{M}_{\odot}\\
		5+1.304(M_{He}-64) &  {\rm if}~140~\mbox{M}_{\odot} \leq M < 260~\mbox{M}_{\odot}\\
		0 & \mbox{otherwise}
	\end{array}
\right.,
\end{equation}
where $M$ is the mass of the zero-age main-sequence star and $M_{He} = (13/24) (M-20)$\msun. The following is the fit to the mass of the remnant \citep{heger2002},
\begin{equation}
\label{eq:remnant}
\frac{M_{rem}(M)}{\mbox{M}_{\odot}} =
\left\{
	\begin{array}{ll}
		1.665 & {\rm if}~11~\mbox{M}_{\odot} \leq M < 20~\mbox{M}_{\odot}\\
		-1.032+0.1349M & {\rm if}~20~\mbox{M}_{\odot} \leq M < 25~\mbox{M}_{\odot}\\
		\sum_{i=0}^4 c_i M^i & {\rm if}~25~\mbox{M}_{\odot} \leq M < 100~\mbox{M}_{\odot}\\
		65 & {\rm if}~100~\mbox{M}_{\odot} \leq M < 140~\mbox{M}_{\odot}\\
		0 & {\rm if}~140~\mbox{M}_{\odot} \leq M < 260~\mbox{M}_{\odot}\\
		-7.761 + 0.8138 M & {\rm if}~260~\mbox{M}_{\odot} < M \leq 500~\mbox{M}_{\odot}
	\end{array}
\right.,
\end{equation}
where $c_0=45.38, c_1=-4.415, c_2=0.1430, c_3=-0.001559$, and $c_4=5.905\times10^{-6}$. Note that a black hole forms when the initial mass is greater than 20\msun. We derive
\begin{equation}
E_{SN}^{\mbox{\small{tot}}}=\int E_{SN}(M) \frac{dN_{pop3}}{d\ln M} d\ln M,
\end{equation}
and
\begin{equation}
M_{rem}^{\mbox{\small{tot}}}=\int M_{rem}(M) \frac{dN_{pop3}}{d\ln M}d\ln M.
\end{equation}
An important parameter of the radiation background is the ratio of the energies per source $\beta \equiv E_{X-ray}/E_{LW}$ \citep{ricotti2016}. We compare the average energies per 100\msun\ in a weak and a strong X-ray backround. We find that total supernova explosion and LW energies per 100\msun\ are comparable, although the two IMFs are different in slope and peak mass. The total supernova explosion energies are $E_{SN}^{\mbox{\small{tot}}} \sim 1.63 \times 10^{52}$ and $2.13 
\times 10^{52}$ erg. In weak X-ray background PISNe are the main energy source while most of the energy comes from hypernovae ($20<M_*<40$\msun) in a strong X-ray background. The energy in the LW band is roughly proportional to the total mass in Pop~III stars, thus for 100\msun\ is $E_{LW}\sim 2.10 \times 10^{53}$ and $1.90 \times 10^{53}$ erg in both backgrounds. Assuming $E_{X-ray}/E_{SN}^{\mbox{\small{tot}}} = 0.006$ \citep[see][]{ricotti2016} the derived values of $\beta$ are: $\beta \sim 4.64\times 10^{-4}$ for weak X-ray irradiation and $\beta \sim 6.74\times 10^{-4}$ for strong X-ray irradiation. Therefore, the IMF of Pop~III stars does not have a strong effect on the spectral energy distribution parameter $\beta$ of the sources. As discussed in Paper~I, however, an X-ray background lowers the critical mass for star formation of halos (hence increases the total number of halos forming Pop~III stars) and reduces the total mass in Pop~III stars per halo. Hence, the X-ray background is nevertheless able to reinforce the feedback loop and promote Pop~III star formation.

A result of interest to LIGO science, is the mass distribution function of Pop~III remnants, shown in Figure~\ref{fig:remmf} for a weak (red histogram) and a strong (blue histogram) X-ray background. The dashed lines show the same distribution function obtained using Monte-Carlo simulations for $10^5$ stars. One notable difference between the distribution functions is the lack of IMBHs with mass $>100$\msun\ and the lack of Pop~III stars exploding as PISNe, for the strong X-ray case. This leads to a higher abundance of low-mass BH remnants with $M < 65$\msun. The Monte-Carlo simulations produce remnants of 158\msun\ and 63.0\msun\ out of 341\msun\ and 102\msun\ Pop~III stars in the two groups. The masses of the remnants per 100\msun\ are 46.3\msun\ and 61.1\msun, respectively.

In future work we will improve on these results considering physical processes neglected or treated with simplifying approximations in this work. In particular, we plan to assess the importance of HD molecular cooling in determining the mass function of Pop~III stars and further investigate the effect of X-ray self-shielding. Most importantly, rather than using the empirical method adopted in this work to estimate the final masses of Pop~III stars, we will use radiative transfer simulations to model the effect of UV radiation feedback from accreting protostar in reducing the accretion rate and stopping the growth of Pop~III stars. 

\section*{Acknowledgements}

We thank Dr. Harley Katz for sharing his version of the RAMSES code with us. All the simulations were performed with the use of Deepthought2 cluster operated by the University of Maryland (http://hpcc.umd.edu). MR acknowledges the support by NASA grant 80NSSC18K0527. KS appreciates the support by the Fellowship of the Japan Society for the Promotion of Science for Research Abroad.




\bibliographystyle{mnras}
\bibliography{mnras} 

\begin{thebibliography}{}
\makeatletter
\relax
\def\mn@urlcharsother{\let\do\@makeother \do\$\do\&\do\#\do\^\do\_\do\%\do\~}
\def\mn@doi{\begingroup\mn@urlcharsother \@ifnextchar [ {\mn@doi@}
  {\mn@doi@[]}}
\def\mn@doi@[#1]#2{\def\@tempa{#1}\ifx\@tempa\@empty \href
  {http://dx.doi.org/#2} {doi:#2}\else \href {http://dx.doi.org/#2} {#1}\fi
  \endgroup}
\def\mn@eprint#1#2{\mn@eprint@#1:#2::\@nil}
\def\mn@eprint@arXiv#1{\href {http://arxiv.org/abs/#1} {{\tt arXiv:#1}}}
\def\mn@eprint@dblp#1{\href {http://dblp.uni-trier.de/rec/bibtex/#1.xml}
  {dblp:#1}}
\def\mn@eprint@#1:#2:#3:#4\@nil{\def\@tempa {#1}\def\@tempb {#2}\def\@tempc
  {#3}\ifx \@tempc \@empty \let \@tempc \@tempb \let \@tempb \@tempa \fi \ifx
  \@tempb \@empty \def\@tempb {arXiv}\fi \@ifundefined
  {mn@eprint@\@tempb}{\@tempb:\@tempc}{\expandafter \expandafter \csname
  mn@eprint@\@tempb\endcsname \expandafter{\@tempc}}}

\bibitem[\protect\citeauthoryear{{Abbott} et~al.,}{{Abbott}
  et~al.}{2016}]{abbott2016}
{Abbott} B.~P.,  et~al., 2016, Phys. Rev. Lett, 116, 161102

\bibitem[\protect\citeauthoryear{{Abe}, {Yajima}, {Khochfar}, {Dalla Vecchia}
  \& {Omukai}}{{Abe} et~al.}{2021}]{Abe2021}
{Abe} M.,  {Yajima} H.,  {Khochfar} S.,  {Dalla Vecchia} C.,   {Omukai} K.,
  2021, arXiv e-prints, \href
  {https://ui.adsabs.harvard.edu/abs/2021arXiv210502612A} {p. arXiv:2105.02612}

\bibitem[\protect\citeauthoryear{{Abel}, {Bryan}  \& {Normal}}{{Abel}
  et~al.}{2002}]{abel2002}
{Abel} T.,  {Bryan} G.~L.,   {Normal} M.~L.,  2002, Science, 295, 93

\bibitem[\protect\citeauthoryear{{Bromm}, {Kudritzki}  \& {Loeb}}{{Bromm}
  et~al.}{2001}]{bromm2001}
{Bromm} V.,  {Kudritzki} R.~P.,   {Loeb} A.,  2001, \apj, 552, 464

\bibitem[\protect\citeauthoryear{{Bromm}, {Coppi}  \& {Larson}}{{Bromm}
  et~al.}{2002}]{bromm2002}
{Bromm} V.,  {Coppi} P.~S.,   {Larson} R.~B.,  2002, \apj, 564, 23

\bibitem[\protect\citeauthoryear{{Chon} \& {Hosokawa}}{{Chon} \&
  {Hosokawa}}{2019}]{Chon2019}
{Chon} S.,  {Hosokawa} T.,  2019, \mn@doi [\mnras] {10.1093/mnras/stz1824},
  \href {https://ui.adsabs.harvard.edu/abs/2019MNRAS.488.2658C} {488, 2658}

\bibitem[\protect\citeauthoryear{{Clark}, {Glover}, {Smith}, Greif, {Klessen}
  \& {Bromm}}{{Clark} et~al.}{2011}]{clark2011}
{Clark} P.~C.,  {Glover} S.~C.~O.,  {Smith} R.~J.,  Greif T.~H.,  {Klessen}
  R.~S.,   {Bromm} V.,  2011, Science, 331, 1040

\bibitem[\protect\citeauthoryear{{Greif}, {Glover}, {Bromm}  \&
  {Klessen}}{{Greif} et~al.}{2010}]{greif2010}
{Greif} T.~H.,  {Glover} S.~C.~O.,  {Bromm} V.,   {Klessen} R.~S.,  2010, \apj,
  716, 510

\bibitem[\protect\citeauthoryear{{Greif}, {Bromm}, {Clark}, {Glover}, {Smith},
  {Klessen}, {Yoshida}  \& {Springel}}{{Greif} et~al.}{2012}]{greif2012}
{Greif} T.~H.,  {Bromm} V.,  {Clark} P.~C.,  {Glover} S.~C.~O.,  {Smith} R.~J.,
   {Klessen} R.~S.,  {Yoshida} N.,   {Springel} V.,  2012, \mnras, 424, 399

\bibitem[\protect\citeauthoryear{{Hahn} \& {Abel}}{{Hahn} \&
  {Abel}}{2011}]{hahn2011}
{Hahn} O.,  {Abel} T.,  2011, \mnras, 415, 2101

\bibitem[\protect\citeauthoryear{{Haiman}, {Abel}  \& {Rees}}{{Haiman}
  et~al.}{2000}]{haiman2000}
{Haiman} Z.,  {Abel} T.,   {Rees} M.~J.,  2000, \apj, 534, 11

\bibitem[\protect\citeauthoryear{{Heath} \& {Nixon}}{{Heath} \&
  {Nixon}}{2020}]{HeatN:2020}
{Heath} R.~M.,  {Nixon} C.~J.,  2020, \mn@doi [\aap]
  {10.1051/0004-6361/202038548}, \href
  {https://ui.adsabs.harvard.edu/abs/2020A&A...641A..64H} {641, A64}

\bibitem[\protect\citeauthoryear{{Heger} \& {Woosely}}{{Heger} \&
  {Woosely}}{2002}]{heger2002}
{Heger} A.,  {Woosely} S.~E.,  2002, \apj, 567, 532

\bibitem[\protect\citeauthoryear{{Hirano} \& {Bromm}}{{Hirano} \&
  {Bromm}}{2017}]{hirano2017}
{Hirano} S.,  {Bromm} V.,  2017, \mnras, 470, 898

\bibitem[\protect\citeauthoryear{{Hirano}, {Hosokawa}, {Yoshida}, {Umeda},
  {Omukai}, {Chiaki}  \& {Yorke}}{{Hirano} et~al.}{2014}]{hirano2014}
{Hirano} S.,  {Hosokawa} T.,  {Yoshida} N.,  {Umeda} H.,  {Omukai} K.,
  {Chiaki} G.,   {Yorke} H.~W.,  2014, \apj, 781, 60

\bibitem[\protect\citeauthoryear{{Hirano}, {Hosokawa}, {Yoshida}, {Omukai}  \&
  {Yorke}}{{Hirano} et~al.}{2015}]{hirano2015}
{Hirano} S.,  {Hosokawa} T.,  {Yoshida} N.,  {Omukai} K.,   {Yorke} H.~W.,
  2015, \mnras, 448, 568

\bibitem[\protect\citeauthoryear{{Hosokawa}, {Omukai}, {Yoshida}  \&
  {Yorke}}{{Hosokawa} et~al.}{2011}]{hosokawa2011}
{Hosokawa} T.,  {Omukai} K.,  {Yoshida} N.,   {Yorke} H.~W.,  2011, Science,
  334, 1250

\bibitem[\protect\citeauthoryear{{Hosokawa}, {Hirano}, {Kuiper}, {Yorke},
  {Omukai}  \& {Yoshida}}{{Hosokawa} et~al.}{2016}]{hosokawa2016}
{Hosokawa} T.,  {Hirano} S.,  {Kuiper} R.,  {Yorke} H.~W.,  {Omukai} K.,
  {Yoshida} N.,  2016, \apj, 824, 119

\bibitem[\protect\citeauthoryear{{Hummel}, {Stacy}, {Jeon}, {Oliveri}  \&
  {Bromm}}{{Hummel} et~al.}{2015}]{hummel2015}
{Hummel} J.~A.,  {Stacy} A.,  {Jeon} M.,  {Oliveri} A.,   {Bromm} V.,  2015,
  \mnras, 453, 4136

\bibitem[\protect\citeauthoryear{{Inayoshi} \& {Omukai}}{{Inayoshi} \&
  {Omukai}}{2011}]{inayoshi2011}
{Inayoshi} K.,  {Omukai} K.,  2011, \mnras, 416, 2748

\bibitem[\protect\citeauthoryear{{Jeon}, {Pawlik}, {Bromm}  \&
  {Milosavljevi'{c}}}{{Jeon} et~al.}{2014}]{jeon2014}
{Jeon} M.,  {Pawlik} A.~H.,  {Bromm} V.,   {Milosavljevi'{c}} M.,  2014,
  \mnras, 440, 3778

\bibitem[\protect\citeauthoryear{{Jeon}, {Bromm}, {Pawlik}  \&
  {Milosavljevi'{c}}}{{Jeon} et~al.}{2015}]{jeon2015}
{Jeon} M.,  {Bromm} V.,  {Pawlik} A.~H.,   {Milosavljevi'{c}} M.,  2015,
  \mnras, 452, 1152

\bibitem[\protect\citeauthoryear{{Kimura}, {Hosokawa}  \& {Sugimura}}{{Kimura}
  et~al.}{2021}]{Kimura2021}
{Kimura} K.,  {Hosokawa} T.,   {Sugimura} K.,  2021, \mn@doi [\apj]
  {10.3847/1538-4357/abe866}, \href
  {https://ui.adsabs.harvard.edu/abs/2021ApJ...911...52K} {911, 52}

\bibitem[\protect\citeauthoryear{{Machacek}, {Bryan}  \& {Abel}}{{Machacek}
  et~al.}{2003}]{machacek2003}
{Machacek} M.~E.,  {Bryan} G.~L.,   {Abel} T.,  2003, \mnras, 338, 273

\bibitem[\protect\citeauthoryear{{Matsukoba}, {Takahashi}, {Sugimura}  \&
  {Omukai}}{{Matsukoba} et~al.}{2019}]{Matsukoba2019}
{Matsukoba} R.,  {Takahashi} S.~Z.,  {Sugimura} K.,   {Omukai} K.,  2019,
  \mn@doi [\mnras] {10.1093/mnras/sty3522}, \href
  {https://ui.adsabs.harvard.edu/abs/2019MNRAS.484.2605M} {484, 2605}

\bibitem[\protect\citeauthoryear{{McKee} \& {Tan}}{{McKee} \&
  {Tan}}{2008}]{mckee2008}
{McKee} C.~F.,  {Tan} J.~C.,  2008, \apj, 681, 771

\bibitem[\protect\citeauthoryear{{Nakauchi}, {Inayoshi}  \&
  {Omukai}}{{Nakauchi} et~al.}{2014}]{nakauchi2014}
{Nakauchi} D.,  {Inayoshi} K.,   {Omukai} K.,  2014, \mnras, 145, 271

\bibitem[\protect\citeauthoryear{{Nomoto}, {Tominaga}, {Umeda}, {Kobayashi}  \&
  {Maeda}}{{Nomoto} et~al.}{2006}]{nomoto2006}
{Nomoto} K.,  {Tominaga} N.,  {Umeda} H.,  {Kobayashi} C.,   {Maeda} K.,  2006,
  Nucl. Phys. A, 777, 424

\bibitem[\protect\citeauthoryear{{Oh}}{{Oh}}{2001}]{oh2001}
{Oh} S.~P.,  2001, \apj, 553, 499

\bibitem[\protect\citeauthoryear{{Omukai}}{{Omukai}}{2001}]{omukai2001}
{Omukai} K.,  2001, \apj, 546, 635

\bibitem[\protect\citeauthoryear{{Omukai} \& {Nishi}}{{Omukai} \&
  {Nishi}}{1998}]{omukai1998}
{Omukai} K.,  {Nishi} R.,  1998, \apj, 508, 141

\bibitem[\protect\citeauthoryear{{Park}, {Ricotti}  \& {Sugimura}}{{Park}
  et~al.}{2021}]{ParkRS:21a}
{Park} J.-W.,  {Ricotti} M.,   {Sugimura} K.,  2021, \mnras, p. submitted

\bibitem[\protect\citeauthoryear{{Planck Collaboration}}{{Planck
  Collaboration}}{2018}]{planck2018}
{Planck Collaboration} 2018, \aap, 641, 6

\bibitem[\protect\citeauthoryear{{Regan}, {Wise}, {O'Shea}  \&
  {Norman}}{{Regan} et~al.}{2020}]{regan2020}
{Regan} J.~A.,  {Wise} J.~H.,  {O'Shea} R.~W.,   {Norman} M.~L.,  2020, \mnras,
  492, 3021

\bibitem[\protect\citeauthoryear{{Ricotti}}{{Ricotti}}{2016}]{ricotti2016}
{Ricotti} M.,  2016, \mnras, 462, 601

\bibitem[\protect\citeauthoryear{{Ricotti} \& {Ostriker}}{{Ricotti} \&
  {Ostriker}}{2004}]{RicottiO:04}
{Ricotti} M.,  {Ostriker} J.~P.,  2004, \mnras, \href
  {http://adsabs.harvard.edu/abs/2004MNRAS.352..547R} {352, 547}

\bibitem[\protect\citeauthoryear{{Ricotti}, {Gnedin}  \& {Shull}}{{Ricotti}
  et~al.}{2002a}]{ricotti2002}
{Ricotti} M.,  {Gnedin} N.~Y.,   {Shull} J.~M.,  2002a, \apj, 575, 33

\bibitem[\protect\citeauthoryear{{Ricotti}, {Gnedin}  \& {Shull}}{{Ricotti}
  et~al.}{2002b}]{RicottiGS:2002b}
{Ricotti} M.,  {Gnedin} N.~Y.,   {Shull} J.~M.,  2002b, \apj, 575, 49

\bibitem[\protect\citeauthoryear{{Ricotti}, {Ostriker}  \& {Gnedin}}{{Ricotti}
  et~al.}{2005}]{RicottiOG:05}
{Ricotti} M.,  {Ostriker} J.~P.,   {Gnedin} N.~Y.,  2005, \mn@doi [\mnras]
  {10.1111/j.1365-2966.2004.08623.x}, \href
  {http://adsabs.harvard.edu/abs/2005MNRAS.357..207R} {357, 207}

\bibitem[\protect\citeauthoryear{{Ricotti}, {Gnedin}  \& {Shull}}{{Ricotti}
  et~al.}{2008}]{RicottiGS:2008}
{Ricotti} M.,  {Gnedin} N.~Y.,   {Shull} J.~M.,  2008, \apj, 685, 21

\bibitem[\protect\citeauthoryear{{Rosdahl}, {Blaizot}, {Aubert}, {Stranex}  \&
  {Teyssier}}{{Rosdahl} et~al.}{2013}]{rosdahl2013}
{Rosdahl} J.,  {Blaizot} J.,  {Aubert} D.,  {Stranex} T.,   {Teyssier} R.,
  2013, \mnras, 436, 2188

\bibitem[\protect\citeauthoryear{{Shull} \& {van Steenberg}}{{Shull} \& {van
  Steenberg}}{1985}]{shull1985}
{Shull} J.~M.,  {van Steenberg} E.~V.,  1985, \apj, 298, 268

\bibitem[\protect\citeauthoryear{{Stacy} \& {Bromm}}{{Stacy} \&
  {Bromm}}{2013}]{stacy2013}
{Stacy} A.,  {Bromm} V.,  2013, \mnras, 433, 1094

\bibitem[\protect\citeauthoryear{{Stacy}, {Bromm}  \& {Lee}}{{Stacy}
  et~al.}{2016}]{stacy2016}
{Stacy} A.,  {Bromm} V.,   {Lee} A.~T.,  2016, \mnras, 462, 1307

\bibitem[\protect\citeauthoryear{{Sugimura}, {Matsumoto}, {Hosokawa}, {Hirano}
  \& {Omukai}}{{Sugimura} et~al.}{2020}]{sugimura2020}
{Sugimura} K.,  {Matsumoto} T.,  {Hosokawa} T.,  {Hirano} S.,   {Omukai} K.,
  2020, \apjl, 892, 14

\bibitem[\protect\citeauthoryear{{Susa}}{{Susa}}{2019}]{Susa2019}
{Susa} H.,  2019, \mn@doi [\apj] {10.3847/1538-4357/ab1b6f}, \href
  {https://ui.adsabs.harvard.edu/abs/2019ApJ...877...99S} {877, 99}

\bibitem[\protect\citeauthoryear{{Susa}, {Kasegawa}  \& {Tominaga}}{{Susa}
  et~al.}{2014}]{susa2014}
{Susa} H.,  {Kasegawa} K.,   {Tominaga} N.,  2014, \apj, 792, 32

\bibitem[\protect\citeauthoryear{{Teyssier}}{{Teyssier}}{2002}]{teyssier2002}
{Teyssier} R.,  2002, \aap, 385, 337

\bibitem[\protect\citeauthoryear{{Toomre}}{{Toomre}}{1964}]{toomre1964}
{Toomre} A.,  1964, \apj, 139, 1217

\bibitem[\protect\citeauthoryear{{Turk}, {Abel}  \& {O'Shea}}{{Turk}
  et~al.}{2009}]{turk2009}
{Turk} M.~J.,  {Abel} T.,   {O'Shea} B.,  2009, Science, 325, 601

\bibitem[\protect\citeauthoryear{{Venkatesan}, {Giroux}  \&
  {Shull}}{{Venkatesan} et~al.}{2001}]{venkatesan2001}
{Venkatesan} A.,  {Giroux} M.~L.,   {Shull} J.~M.,  2001, \apj, 563, 1

\bibitem[\protect\citeauthoryear{{Whalen}, {Smidt}, {Even}, {Woosley}, {Heger},
  {Stiavelli}  \& {Fryer}}{{Whalen} et~al.}{2014}]{whalen2014}
{Whalen} D.~J.,  {Smidt} J.,  {Even} W.,  {Woosley} S.~E.,  {Heger} A.,
  {Stiavelli} M.,   {Fryer} C.~L.,  2014, \apj, 781, 106

\bibitem[\protect\citeauthoryear{{Wise} \& {Abel}}{{Wise} \&
  {Abel}}{2008}]{wise2008}
{Wise} J.~H.,  {Abel} T.,  2008, \apj, 685, 40

\bibitem[\protect\citeauthoryear{{Wise}, {Turk}, {Norman}  \& {Abel}}{{Wise}
  et~al.}{2012}]{wise2012}
{Wise} J.~H.,  {Turk} M.~J.,  {Norman} M.~L.,   {Abel} T.,  2012, \apj, 745, 50

\bibitem[\protect\citeauthoryear{{Wolcott-Green} \& {Haiman}}{{Wolcott-Green}
  \& {Haiman}}{2019}]{wolcottgreen2019}
{Wolcott-Green} J.,  {Haiman} Z.,  2019, \mnras, 484, 2467

\bibitem[\protect\citeauthoryear{{Xu}, {Ahn}, {Wise}, {Norman}  \&
  {O'Shea}}{{Xu} et~al.}{2014}]{xu2014}
{Xu} H.,  {Ahn} K.,  {Wise} J.~H.,  {Norman} M.~L.,   {O'Shea} B.~W.,  2014,
  \apj, 791, 110

\bibitem[\protect\citeauthoryear{{Xu}, {Ahn}, {Normal}, {Wise}  \&
  {O'Shea}}{{Xu} et~al.}{2016}]{xu2016}
{Xu} H.,  {Ahn} K.,  {Normal} M.~L.,  {Wise} J.~H.,   {O'Shea} B.~W.~O.,  2016,
  \apjl, 831, L5

\bibitem[\protect\citeauthoryear{{Yoshida}, {Oh}, {Kitayama}  \&
  {Hernquist}}{{Yoshida} et~al.}{2007}]{yoshida2007}
{Yoshida} N.,  {Oh} S.~P.,  {Kitayama} T.,   {Hernquist} L.,  2007, \apj, 663,
  687

\bibitem[\protect\citeauthoryear{{Yoshida}, {Omukai}  \& {Hernquist}}{{Yoshida}
  et~al.}{2008}]{yoshida2008}
{Yoshida} N.,  {Omukai} K.,   {Hernquist} L.,  2008, Science, \href
  {http://adsabs.harvard.edu/abs/2008Sci...321..669Y} {321, 669}

\makeatother
\end{thebibliography}








\bsp	
\label{lastpage}
\end{document}